\documentclass[graybox]{svmult}

\usepackage{mathptmx}       % selects Times Roman as basic font
\usepackage{helvet}         % selects Helvetica as sans-serif font
\usepackage{courier}        % selects Courier as typewriter font
\usepackage{type1cm}        % activate if the above 3 fonts are
\usepackage{makeidx}         % allows index generation
\usepackage{graphicx}        % standard LaTeX graphics tool
\usepackage{multicol}        % used for the two-column index
\usepackage[bottom]{footmisc}% places footnotes at page bottom

\newcommand{\bea}{\begin{eqnarray}}
\newcommand{\eea}{\end{eqnarray}}
\newcommand{\bi}{\begin{itemize}}
\newcommand{\ei}{\end{itemize}}
\def\Tdot#1{{{#1}^{\hbox{.}}}}
\def\Tddot#1{{{#1}^{\hbox{..}}}}

\newcommand{\dn}[2]{{\mathrm{d}^{{#1}}{{#2}}}}

\def\s{\sigma}

\def\h{{\cal H}}

\def\d{\delta}
\def\B{{\bar B}}
\def\E{{\bar E}}
\def\V{{\cal V}}
\def\P{{\cal P}}
\def\R{{\cal R}}

\def\x{{\vec x}}

\def\mP{M_P}

\def\bfphi{{\bf \phi}}

\def\bx{{\bf x}}
\def\bk{{\bf k}}
\def\PP{{\cal P}}

\def\beq{\begin{equation}}
\def\eeq{\end{equation}}

\def\h{{\cal H}}

\def\d{\delta}
\def\B{{\bar B}}
\def\E{{\bar E}}
\def\V{{\cal V}}
\def\P{{\cal P}}
\def\R{{\cal R}}

\def\x{{\vec x}}

\def\mP{M_P}  % reduced Planck mass=1/\sqrt{8\pi G}
\newcommand{\square}{\kern1pt\vbox{\hrule height
1.2pt\hbox{\vrule width 1.2pt\hskip 3pt
\vbox{\vskip 6pt}\hskip 3pt\vrule width 0.6pt}\hrule
height 0.6pt}\kern1pt}

\makeindex             % used for the subject index
\begin{document}

\title*{Lectures on inflation and cosmological perturbations}
\titlerunning{Inflation and cosmological perturbations}
\index{inflation}
\author{David Langlois}
\institute{APC (CNRS-Universit\'e Paris 7),\\ 
10, rue Alice Domon et L\'eonie Duquet, 75205 Paris Cedex 13, France}
\maketitle

\abstract{\\
The purpose of these lectures is  to  give a  pedagogical introduction to  inflation and  the production of the 
primordial perturbations, as well as a review of some of the latest developments in this domain. 
\\
After a short introduction, we review the main principles of the Hot Big Bang model, as well as its limitations. These deficiencies provide the motivation for the study of a cosmological phase of accelerated expansion, called inflation, which can be  induced by a slow-rolling scalar field. A  few illustrative models  are presented. We then turn to the analysis of cosmological perturbations, and explain how the vacuum quantum fluctuations are amplified during an inflationary phase. The next step consists in relating the perturbations generated during inflation to the perturbations of the cosmological fluid in the standard radiation dominated phase. One can thus confront the predictions of inflationary models with  cosmological observations, such as the measurements  of the Cosmic Microwave Background or the large-scale structure surveys. The present constraints on inflationary models are discussed. 
\\
The final part of these lectures gives a review of more general models of inflation, involving multiple fields or non standard kinetic terms. Although more complicated, these models are usually  motivated by high energy physics and they  can lead to specific signatures that are not expected in the simplest models of inflation. After introducing a very general formalism to describe perturbations in multi-field models with arbitrary kinetic terms, several interesting cases are presented. We also stress the role of  entropy perturbations in the context of multi-field models. Finally,  we discuss in detail  the non-Gaussianities of the primordial perturbations and some models that could produce a detectable level of non-Gaussianities. 
 }
\rm

%%%%%%%%%%%%%%
\section{Introduction}
%%%%%%%%%%%%%%

 Inflation is today the main theoretical framework that describes  the early Universe and that can account for the present observational data. In thirty years of existence,  inflation has survived, in contrast with earlier competitors,  the tremendous improvement of cosmological data.  In particular, the fluctuations of the Cosmic Microwave Background (CMB) had not yet been measured when inflation was invented, whereas they give us today  a remarkable   picture of  the cosmological perturbations in the early Universe.  In the future, one can hope that more precise  measurements of the primordial cosmological perturbations will allow us to go one step further in the confrontation of   inflation models with data, and especially to discriminate between the many different possible realizations of inflation.

The purpose of these lectures is two-fold. The first goal is  to explain,  in a simple way  and starting from first principles as much as possible, the conceptual basis of inflation and the elementary steps to calculate the cosmological perturbations predicted by  the simplest models. 
The second objective of these lectures is to give an overview of  the latest developments on inflation, in particular   the study of more general models of inflation involving several scalar fields or non-standard kinetic terms. Although more complicated, these models can give very specific signatures  in  the primordial cosmological perturbations, in particular non-Gaussianities and isocurvature perturbations. 

 There is a huge literature on inflation and these lectures cover only a few topics, with a list of references  
  that is far from exhaustive. More details and  more references can be found in several textbooks (see e.g. \cite{linde, ll,mukhanov}) and many reviews (including for instance  \cite{Lidsey:1995np,lr,cargese,Riotto:2002yw,Bassett:2005xm}; more specialized reviews will  also be mentioned in the text). A novel feature of these lectures is to introduce the latest  methods used for  the computation of perturbations. They have the advantage to be easily extendible to the study of non-linear perturbations, which has recently become an extremely active topic. 
  
 The outline of these lectures is the following. In the next section, we recall the basic elements of the Hot Big Bang model and discuss its limitations, which motivate inflation. Homogeneous inflation is introduced in Section 3. In Section 4, we turn to the theory of linear cosmological perturbations and explain how they are generated during an inflationary phase. The following section, Section 5, is devoted to the link between primordial perturbations and present cosmology, and thus to the confrontation of inflation models with the data. In Section 6, more general models of inflation are considered, with a discussion of several specific scenarios, which have attracted a lot of attention recently. Section 7 is devoted to the primordial non-Gaussianities and we conclude in the last section.

\section{The hot Big Bang model}
Modern cosmology is based on the theory of general relativity, according to which our Universe is described by a four-dimensional 
geometry $g_{\mu\nu}$ that  satisfies Einstein's equations
\beq
\label{einstein}
G_{\mu\nu}\equiv R_{\mu\nu}-\frac12 R \, g_{\mu\nu}=8\pi G \, T_{\mu\nu},
\eeq
where $R_{\mu\nu}$ is the Ricci tensor, $R\equiv g^{\mu\nu}R_{\mu\nu}$ the scalar curvature and $T_{\mu\nu}$ the energy-momentum tensor that describes the matter distribution. 

\subsection{The Friedmann equations}

One of the main assumptions of cosmology, which has  
been confirmed by observations so far,  
is to consider, as a first approximation, the universe as 
being homogeneous and 
isotropic. Note that these symmetries define  implicitly a particular 
``slicing'' of spacetime, in which the space-like hypersurfaces 
are homogeneous and isotropic. A different slicing of the {\it same}
spacetime would give   space-like hypersurfaces that are not 
homogeneous and isotropic.

Homogeneity and isotropy turn out to be very restrictive and the only geometries 
compatible with these requirements   are the FLRW (Friedmann-Lema\^itre-Robertson-Walker) spacetimes, with  metric
\beq
ds^2=-dt^2+a^2(t)\left[{dr^2\over{1-\kappa r^2}}+
r^2\left(d\theta^2+\sin^2\theta d\phi^2\right)\right],
\label{RW}
\eeq
where $\kappa=0,1,-1$ determines the curvature of spatial 
hypersurfaces: respectively flat, elliptic or hyperbolic.
Moreover, the  matter content compatible with  homogeneity
and isotropy is necessarily characterized by an energy-momentum tensor of the 
form 
\beq
\label{T}
T^{\mu}_{\  \nu}={\rm Diag}\left[-\rho(t), P(t),  P(t), P(t)\right]\, 
\eeq
where $\rho$ corresponds to an energy density and $P$ to a pressure.

Substituting the  metric (\ref{RW})  and the energy-momentum tensor (\ref{T}) 
into Einstein's equations (\ref{einstein}) gives the  
Friedmann equations\index{Friedmann equations},
\begin{eqnarray}
\left({\dot a\over a}\right)^2 &=& {8\pi G\rho\over 3}- {\kappa\over a^2},
\label{friedmann}
\\
{\ddot a\over a} &=& -{4\pi G\over 3}\left(\rho+3 P\right)\, ,
\label{fried2}
\end{eqnarray}
which govern the time evolution of the scale factor $a(t)$. 

An immediate consequence of the two above equations is the {\it continuity 
equation}
\beq
\dot \rho+3H\left(\rho+P\right)=0,
\label{conservation}
\eeq
where $H\equiv \dot a/a$ is the {\it Hubble parameter}.
The continuity equation 
 can  also be obtained directly from the energy-momentum conservation\index{energy-momentum conservation}
\beq
 \nabla_\mu
T^{\mu}_{\ \nu}=0,
\eeq
where $\nabla$ denotes the covariant derivative associated with the metric $g_{\mu\nu}$.

The cosmological evolution can be determined once the equation of state for the matter is specified. 
 Let us assume  here $P=w\rho$ with
$w$ constant, which includes the two main types of matter that play an 
important r\^ole in cosmology, namely non relativistic matter ($w\simeq 0$) and a gas of relativistic particles ($w=1/3$).
The conservation equation  (\ref{conservation}) 
can be integrated to give
\beq
\rho\propto a^{-3(1+w)}.
\eeq
Substituting into  (\ref{friedmann}), one finds,  for  $\kappa=0$,  
\beq
a(t)\propto t^{\frac{2}{3(1+w)}},
\eeq
which thus gives the evolution $a(t)\propto t^{1/2}$ for relativistic matter and $a(t)\propto t^{2/3}$ for non-relativistic matter. Note that a different cosmological evolution, governed by modified Friedmann's equations, can be envisaged in the primordial Universe,  as for example in the context of brane cosmology (see e.g. \cite{Langlois:2005nd}), but this possibility will not be discussed in these notes.

The present cosmological observations seem to indicate that our Universe is currently accelerating. The simplest way to account for this acceleration is to assume the presence of a {\it cosmological constant}  $\Lambda$ in Einstein's equations, i.e. an additional term $\Lambda g_{\mu\nu}$ on the left-hand side of (\ref{einstein}). By moving this  term on the right hand side of Einstein's equations it can also be interpreted   as an energy-momentum tensor with equation of state $P=-\rho$, where $\rho$ is time-independent. This leads, for $\kappa=0$ and without any other matter, to 
an exponential evolution of the scale
factor
\beq
a(t)\propto \exp(Ht).
\eeq

In our universe, several species with different equations of state coexist, and it has become customary to characterize their relative contributions by the dimensionless parameters
\beq
\Omega_{(i)}\equiv \frac{8\pi G\rho_0^{(i)}}{3H_0^2},
\eeq
where the $\rho_0^{(i)}$ denote the present energy densities of the various species, and $H_0$ is the present Hubble parameter.  The first Friedmann equation (\ref{friedmann}), evaluated at the present time,  implies 
\beq
\Omega_0=\sum_i\Omega_{(i)}=1+\frac{\kappa}{a_0^2H_0^2}.
\eeq
One can infer from  present observations the following parameters:  $\Omega_m\simeq 0.3$ for non-relativistic matter (which includes a small baryonic component $\Omega_b\simeq 0.05$), $\Omega_\Lambda\simeq 0.7$ for a ``dark energy" component (compatible with a cosmological constant), $\Omega_{\gamma}\simeq 5\times 10^{-5}$ for the photons, and a total $\Omega_0$ close to $1$, i.e. no detectable deviation from flatness.

\subsection{The shortcomings of the standard Big Bang model}

The standard Big Bang model has encountered remarkable successes, in 
particular with primordial nucleosynthesis  and the CMB, 
and it remains today a cornerstone in our understanding of the present 
and past universe.
However, a few intriguing facts  remain unexplained in the strict 
scenario of the Hot Big Bang model and seem to necessitate a larger 
framework. We review now the main problems:

\begin{itemize}

\item Homogeneity problem\index{homogeneity problem}

A first question is why the approximation of homogeneity and isotropy 
turns out to be so good. Indeed, inhomogeneities are 
unstable, because of gravitation, and they tend to grow with time. 
It can be verified for instance with the CMB that inhomogeneities were 
much smaller at the last scattering epoch than today. One thus expects that 
these homogeneities were still smaller further back in time.
How to explain a universe so smooth in its past ?
\\

\item Flatness problem\index{flatness problem}

Another puzzle lies in the (spatial) flatness of our universe. Indeed, the
first Friedmann equation, Eq.~(\ref{friedmann}),  implies
\beq
\Omega-1\equiv {8\pi G \rho\over 3 H^2}-1={\kappa\over a^2 H^2}.
\eeq
In standard cosmology, the scale factor behaves like
$a\sim t^p$ with $p<1$  ($p=1/2$ for radiation and $p=2/3$ for 
non-relativistic  matter).
 As a consequence,  $(aH)^{-2}$  grows with time and 
  $|\Omega-1|$ must thus  diverge with time. Therefore, in the 
context of the standard model, the quasi-flatness observed today 
requires an extreme fine-tuning of $\Omega$ near $1$ in the early 
 universe. 
\\

\item Horizon problem\index{horizon problem}

One of the most fundamental problems in standard cosmology is certainly
the {\it horizon problem}.
The (particle) {\it horizon} is the maximal distance that can be covered 
by a light ray. 
For a light-like radial trajectory, 
 $dr=a(t) dt$ and   the horizon is thus given by 
\beq
d_{H}(t)= a(t)\int_{t_i}^t{dt'\over a(t')}=a(t){t^{1-q}-t_i^{1-q}\over 1-q}, 
\eeq
where the last equality is obtained by assuming $a(t)\sim t^q$ and $t_i$ is
some initial time.

In standard cosmology ($q<1$), the integral converges in the limit 
 $t_i=0$ and the horizon has a finite size, of the order of the 
so-called Hubble radius\index{Hubble radius} $H^{-1}$:
\beq
d_H(t)={q\over 1-q}H^{-1}.
\eeq 
 
It also useful to consider 
 the {\it comoving Hubble radius}, $(aH)^{-1}$, which represents
the fraction of comoving space in causal contact.
One finds that it  {\it grows} with time, which means that the 
{\it fraction of the universe in causal contact increases with time} in 
the context of standard cosmology.
But the CMB tells us that the Universe was quasi-homogeneous 
at the time of last scattering on a scale encompassing 
 many regions a priori causally independent.
How to explain this ? 
\\

\end{itemize}

A solution to the horizon problem and to the other puzzles 
is provided by the inflationary scenario, which 
we will examine in the next section. The basic idea
 is to ``decouple'' the causal size from the Hubble radius, so that the real size of the horizon region 
 in the standard  radiation dominated era  is much larger than the  Hubble radius. 
 Such a situation occurs if  the comoving Hubble radius {\it decreases} sufficiently in the very 
early universe. 
The corresponding condition is 
\beq
\ddot a>0,
\eeq
i.e.  the Universe undergoes a {\it phase of acceleration}. 

%%%%%%%%%%%%%%%%%%%%%%
%%%%%%%%%%%%%%%%%%%%%%
\section{Inflation}
%%%%%%%%%%%%%%%%%%%%%%
%%%%%%%%%%%%%%%%%%%%%%
The broadest definition of inflation is that it corresponds to a phase
of acceleration of the universe,
\beq
\ddot a>0.
\eeq
In this  sense, the current cosmological observations, if correctly
interpreted, mean that our present universe is undergoing 
 an inflationary phase. 
It is worth noting that many of the models suggested for inflation have been 
adapted to account for the present acceleration. 
 We are however interested here in an inflationary 
phase taking place in the {\it early} universe, thus characterized by  very different energy 
scales. Another difference is that inflation in the early universe {\it must end}  to leave room to the standard
radiation dominated cosmological phase.

Cosmological acceleration requires, according to the second Friedmann equation, Eq.~(\ref{fried2}),   an 
 equation of state satisfying  
\beq
P<-{1\over 3}\rho,
\eeq
condition which looks at first view   rather exotic.

A very simple example giving such an equation of state is 
a cosmological constant, corresponding to a cosmological fluid
with the equation of state
\beq
P=-\rho.
\eeq
However, a strict cosmological constant leads to 
exponential inflation {\it forever} which cannot be followed by 
a radiation era.
Another possibility is a scalar field, which we now  discuss  in some 
details. 

\subsection{Cosmological scalar fields}
\index{scalar field}

The dynamics  of a scalar field minimally coupled to gravity is governed by the 
action 
\beq
\label{action_scalar_field}
S_\phi=\int d^4x\sqrt{-g}\left(-{1\over 2}\partial^\mu\phi\partial_\mu\phi
-V(\phi)\right)\, ,
\eeq
where $g\equiv {\rm det} (g_{\mu\nu})$ and $V(\phi)$ is the potential of the scalar field.
The corresponding energy-momentum tensor, obtained by varying the action (\ref{action_scalar_field}) with respect to the metric,  is given by  
\beq
T_{\mu\nu}=\partial_\mu\phi\partial_\nu\phi-g_{\mu\nu}
\left({1\over 2}\partial^\sigma\phi\partial_\sigma\phi
+V(\phi)\right).
\label{Tscalarfield}
\eeq
In the homogeneous and isotropic geometry (\ref{RW}),  
the energy-momentum tensor is of the perfect fluid form, 
with the energy density
\beq
\rho=-T_0^0={1\over 2}\dot\phi^2+V(\phi),
\eeq
where one recognizes the sum of a kinetic energy and 
of a potential energy, and the pressure
\beq
P={1\over 2}\dot\phi^2-V(\phi).
\eeq
The equation of motion for the scalar field is the Klein-Gordon 
equation, obtained by taking the variation of the above 
action (\ref{action_scalar_field}) 
with respect to the scalar field, 
\index{Klein-Gordon equation}
\beq
\nabla^\mu\nabla_\mu\phi=\frac{dV}{d\phi},
\eeq
which reduces to
\beq
\ddot\phi+3H\dot \phi+V'=0
\eeq
in  a homogeneous and isotropic universe.

The system of equations governing the dynamics of the scalar field 
and of the cosmological geometry  is thus given by
\begin{eqnarray}
& &H^2={8\pi G\over 3}\left({1\over 2}\dot\phi^2+V(\phi)\right), 
\label{e1}\\
& &\ddot\phi+3H\dot \phi+V'=0, 
\label{e2}\\
& & \dot H=-4\pi G\dot\phi^2.
\label{e3}
\end{eqnarray}
The last equation can be derived from the first two and is therefore
redundant. 

%%%%%%%%%%%%%%%%%%%%%%%%%%%%%%%%%%
\subsection{The slow-roll regime}
%%%%%%%%%%%%%%%%%%%%%%%%%%%%%%%%%
\index{slow-roll}
The dynamical system (\ref{e1}-\ref{e3}) does not always give an accelerated
expansion but it does so in the so-called {\it slow-roll regime} when 
the potential energy of the scalar field dominates over its kinetic 
energy. 

More specifically, 
 the slow-roll approximation consists in neglecting 
the kinetic energy of the scalar field, $\dot \phi^2$, in (\ref{e1}) 
and its acceleration, $\ddot\phi$, in the  Klein-Gordon equation 
(\ref{e2}). One then gets the simplified system
\begin{eqnarray}
& &H^2\simeq{8\pi G\over 3} V, 
\label{sr1}\\
& &3H\dot \phi+V'\simeq 0. 
\label{sr2}
\end{eqnarray}
Let us now examine in which regime this  approximation is valid. 
From
(\ref{sr2}), 
the velocity of the scalar field is given by
\beq
\dot\phi\simeq -{V'\over 3H}.
\label{phisr}
\eeq
Substituting this relation into the  condition $\dot \phi^2/2 \ll V$ yields 
the requirement
\beq
\epsilon_V\equiv {\mP^2\over 2}\left({V'\over V}\right)^2 \ll 1,
\label{epsilon}
\eeq
where we have introduced the {\it reduced Planck mass}
\beq
M_P\equiv {1\over \sqrt{8\pi G}}.
\eeq
Alternatively, one can use the parameter
\beq
\epsilon\equiv -\frac{\dot H}{H^2},
\eeq
which coincides with $\epsilon_V$ at leading order in slow-roll, since $\epsilon=\dot\phi^2/(2 M_P^2 H^2)$.

Similarly, $\ddot \phi\ll V'$ implies, after using the time derivative of (\ref{phisr}) and (\ref{sr1}), the condition 
\beq
\eta_V\equiv \mP^2 {V''\over V}\ll 1.
\label{eta}
\eeq
In summary, the slow-roll approximation is valid when the 
 conditions $\epsilon_V, \eta_V \ll 1$ are satisfied by the potential, which means 
that the slope and the curvature of the potential, in Planck units, 
must be sufficiently small.

%%%%%%%%%%%%%%%%%%%%%%%%%%%%%%%
\subsection{Number of e-folds}
%%%%%%%%%%%%%%%%%%%%%%%%%%%%%%%
\index{number of e-folds}

Inflation must last long enough, in order to solve the problems of the Hot Big Bang model. 
To investigate this question, one usally introduces 
the {\it number of e-folds before the end of inflation}, denoted $N$,
and simply defined by 
\beq
N=\ln {a_{end}\over a},
\eeq
where $a_{end}$ is the value of the scale factor at the end of inflation
and $a$ is a fiducial value for the scale factor during inflation.
By definition, $N$ {\it decreases} during the inflationary phase 
and reaches zero at its end. 

In the slow-roll approximation, it is possible to express 
$N$ as a function of the scalar field. Since 
 $dN=-d\ln a=-H dt=-(H/\dot\phi) d\phi$, one easily finds, using
(\ref{phisr}) and (\ref{sr1}), that 
\beq
N(\phi)\simeq \int_\phi^{\phi_{end}}{V\over M_P^2 V'}d\phi.
\eeq
Given an explicit potential $V(\phi)$, one can in principle integrate 
the above expression to obtain $N$ in terms of $\phi$. This will be 
illustrated below for  some specific models.

Let us now discuss  the link  between $N$ and the present 
cosmological scales. If one considers a given scale characterized by 
its comoving wavenumber $k=2\pi/\lambda$,
this scale crossed out the Hubble radius, 
during inflation, at an  instant $t_*(k)$ defined by 
\beq
k=a(t_*) H(t_*).
\eeq
To get a rough estimate  of the number of e-foldings of 
inflation that are needed 
to solve the horizon problem, let us first ignore the transition 
from a radiation era to a matter era and assume for simplicity that 
the inflationary phase was followed instantaneously 
by a radiation phase that has lasted   until now.
During the radiation phase, the comoving Hubble radius  $(aH)^{-1}$ 
increases like $a$. In order to solve the horizon problem, the increase
of the comoving Hubble radius during the standard evolution 
must be compensated by {\it at least}
a decrease of the same amount during inflation. 
Since the comoving Hubble radius 
roughly scales like $a^{-1}$ during inflation, the minimum amount 
of inflation is simply 
given by the number of e-folds between the end of inflation 
and today  
\beq
\ln(a_0/a_{end}) = 
\ln(T_{end}/T_0)\sim \ln(10^{29}(T_{end}/ 10^{16} {\rm GeV})),
\eeq
 i.e. 
around 60 e-folds for a temperature $T\sim 10^{16} {\rm Gev}$ 
at the beginning of the radiation era. As we will see later, this 
energy scale is typical of inflation  in the simplest models.

%%%%%%%%%%%%%%%%%%%%%%%%%%%%%%%%%%%%%%%%%%%%%%%%%%%%
\begin{figure}
\begin{center}
\includegraphics[width=4.5in]{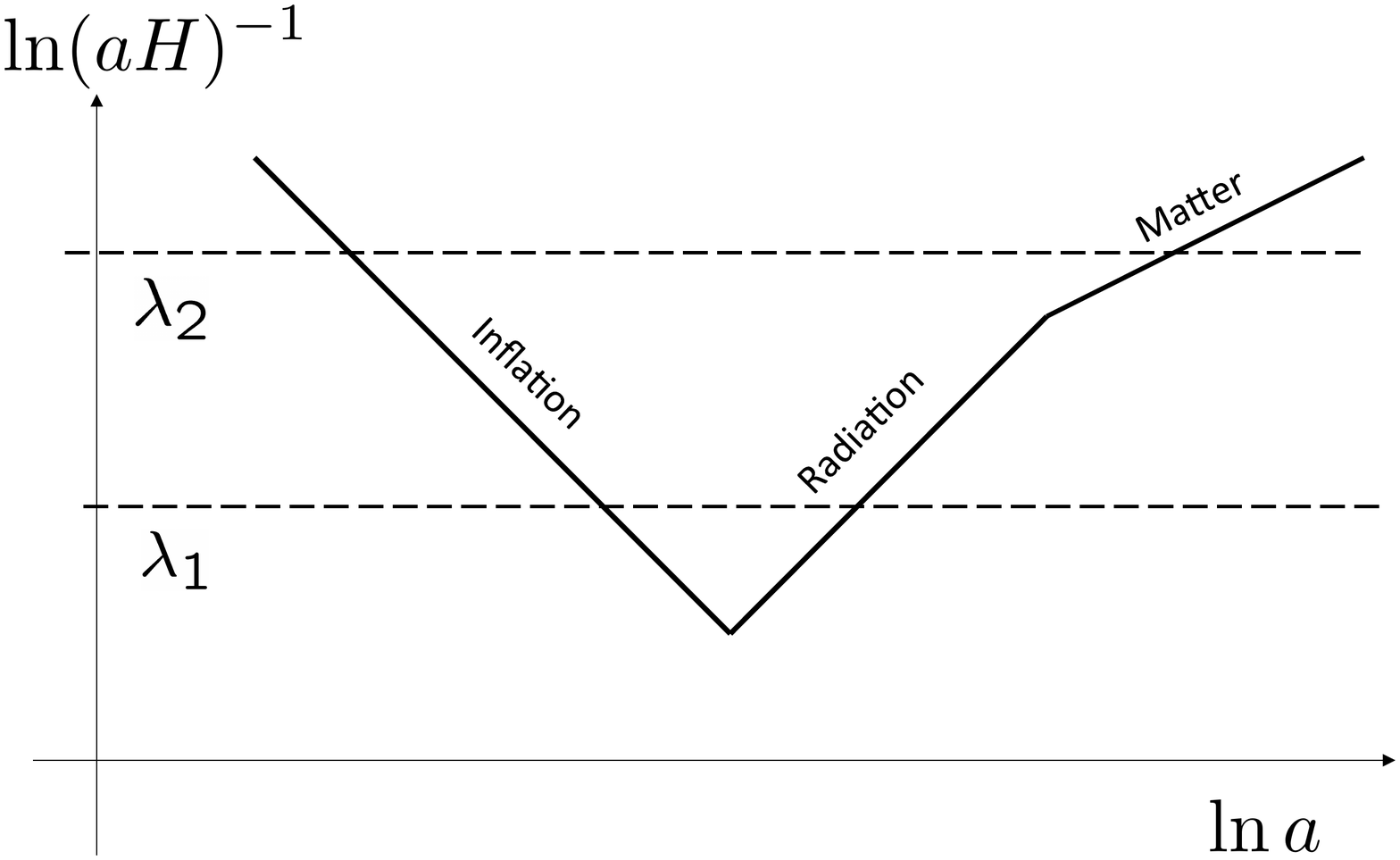}
\end{center}
\caption{Evolution of the comoving Hubble radius $\lambda_H=(aH)^{-1}$,  during inflation, radiation dominated era and matter 
dominated era. The horizontal dashed lines correspond to two different comoving lengthscales: the larger scales cross out the Hubble radius {\it earlier} during inflation and reenter the Hubble radius {\it later} in the standard cosmological era.}
\end{figure}
%%%%%%%%%%%%%%%%%%%%%%%%%%%%%%%%%%%%%%%%%%%%%%%%%%%%

This determines roughly the number of e-folds $N(k_0)$ 
between the moment when the scale corresponding to our present 
Hubble radius $k_0=a_0H_0$ exited the Hubble radius during inflation and 
the end of inflation.
 The other lengthscales  of cosmological interest are {\it smaller} than 
$k_0^{-1}$ and therefore exited the Hubble radius during inflation 
{\it after}  the scale $k_0$, whereas they entered the Hubble 
radius during the standard cosmological phase (either in the radiation era
for the smaller scales or in the matter era for the larger scales) 
{\it before} the scale $k_0$  (see Fig. 1).

A more detailed calculation, which distinguishes between the energy
scales at the end of inflation and after the reheating, gives
for the number of e-folds between the exit of the mode $k$ and the end
of inflation (see e.g. \cite{ll,Liddle:2003as})
\beq
N(k)\simeq 62 -\ln{k\over a_0 H_0}+\ln{V_k^{1/4}\over 10^{16}{\rm GeV}}
+ \ln{V_k^{1/4}\over V_{\rm end}^{1/4}}
+{1\over 3}\ln{\rho_{\rm reh}^{1/4}\over 
 V_{\rm end}^{1/4}}.
\eeq
Since the
 smallest  scale of cosmological relevance 
is  of the order of  $1$ Mpc, the range of cosmological 
scales  covers about $9$ e-folds.

The above number of e-folds is altered if one changes the thermal 
history of the universe between inflation and the present time 
 by including for instance
a period of so-called thermal inflation.

%%%%%%%%%%%%%%%%%%%%%%%%%%%%%%%%%%
\subsection{A few examples}
%%%%%%%%%%%%%%%%%%%%%%%%%%%%%%%%%%
It is now time to illustrate all the points  discussed above 
with some specific potentials. 

\subsubsection{Power-law potential}
\index{power-law potential}
We consider first the case of power-law monomial 
potentials, of the form 
\beq
\label{pot_power-law}
V(\phi)=\lambda\phi^p,
\eeq
which have been abundantly studied in the literature. In particular, 
the above potentials include the case of a free massive scalar field, 
$V(\phi)=m^2\phi/2$.

The slow-roll  parameters are given by
\beq
\label{SR_param_power-law}
\epsilon=\frac{p^2 M_P}{2\phi^2}, \qquad \eta=p(p-1) \frac{M_P^2}{\phi^2}.
\eeq
The slow-roll conditions
  $\epsilon\ll 1$ and  $\eta \ll 1$ 
thus imply   
\beq 
\phi \gg p \  \mP,
\label{sr_pl}
\eeq
which means that the scalar field amplitude must be above the Planck mass
during inflation. 

After substituting the potential (\ref{pot_power-law}) into the 
slow-roll equations of motion (\ref{sr1}-\ref{sr2}), 
one can integrate them explicitly to get 
\beq
\phi^{2-{p\over 2}}-  \phi_i^{2-{p\over 2}}
 =-\frac{2p}{4-p} \, \sqrt{\frac{\lambda}{3} }\, 
\mP\left(t-t_i\right)
\eeq
for $p\neq 4$ and
\beq
\phi=\phi_i\exp\left[-4\sqrt{\lambda\over 3}\,  \mP (t-t_i)\right]
\eeq
for $p=4$.

One can also express the scale factor as a function of the scalar field
[and thus as a function of time by substituting the above expression for
 $\phi(t)$] by using  $d\ln a/ d\phi=H/\dot \phi
\simeq-\phi/(p M_P^2)$. One finds 
\beq
a=a_{end}\exp\left[-{\left(\phi^2-\phi_{end}^2\right)\over 2p\,  M_P^2}\right].
\eeq
Defining the end of inflation by $\epsilon=1$, one gets 
$\phi_{end}=p \, \mP/\sqrt{2}$ and the number of e-folds is thus given by
\beq
\label{N_phi}
N(\phi)\simeq {\phi^2\over 2p \mP^2}-{p\over 4}.
\eeq
This can be inverted, so that
\beq
\phi(N)\simeq \sqrt{2Np}\, \mP,
\eeq
where we have ignored the second term of the right hand side of 
(\ref{N_phi}), consistently  with the condition (\ref{sr_pl}).

%%%%%%%%%%%%%%%%%%%%%%%%%%%%%%%%%%%%
\subsubsection{Exponential potential}
%%%%%%%%%%%%%%%%%%%%%%%%%%%%%%%%%%%%
\index{exponential potential}
Cosmological scalar fields with a potential of the form
\beq
\label{potential_exponential}
V=V_0 \exp \left(-\sqrt{2\over q}\, {\phi\over M_P}\right),
\eeq
admit an {\it exact} solution (i.e. valid beyond 
the slow-roll approximation) of the system (\ref{e1}-\ref{e3}), with a
power-law   scale factor, i.e.
\beq
a(t)\propto t^q.
\eeq
The evolution of the scalar field is given by the expression
\beq 
\phi(t)={\sqrt{2q}\, M_P}\ln\left[\sqrt{V_0\over q(3q-1)} {t\over M_P}\right].
\eeq
Note that  one recovers the slow-roll approximation in the limit 
 $q\gg 1$, since the slow-roll parameters are given by 
 \beq
 \label{SR_param_exponential}
 \epsilon_V=\frac{1}{q}\, \qquad  
\eta_V=\frac{2}{q}\, .
\eeq

\subsubsection{Hybrid inflation}
\index{hybrid inflation}
In this type of model,  the potential contains a constant piece in addition to a power-law potential, the simplest example being
\beq
\label{hybrid}
V(\phi)=V_0+{1\over 2}m^2\phi^2.
\eeq
In fact, the full model relies on the presence 
of two scalar fields, where one plays the traditional r\^ole of the inflaton, 
while the other is  necessary to end inflation.
In the original model of hybrid 
inflation \cite{hybrid}, one starts from  the potential 
\beq
V(\phi, \psi)={1\over 2}m^2\phi^2+{1\over 2}\lambda' \psi^2\phi^2+
{1\over 4}\lambda\left(M^2-\psi^2\right)^2.
\eeq
For values of the field $\phi$ larger than the critical value 
 $\phi_c=\lambda M^2/\lambda'$, the potential for $\psi$ has its minimum 
at $\psi=0$. This is the case during inflation: $\psi$ is thus
trapped in this minimum $\psi=0$, so that the effective potential 
for the scalar field $\phi$, which plays the r\^ole of the inflaton, is 
given by (\ref{hybrid}) 
with $V_0=\lambda M^4/4$.
 During the inflationary phase, the field $\phi$ slow-rolls until it reaches 
the critical value  $\phi_c$. The shape of the potential for $\psi$ is then 
modified and new minima appear in  $\psi=\pm M$.  $\psi$ will thus roll down
into one of these new minima and, as a consequence, inflation will end. 

During the inflationary phase, the slow-roll parameters are given by 
\beq
\epsilon=\frac{m^2 M_P^2 \tilde\phi^2}{V_0 (1+\tilde\phi^2)^2}, \qquad 
\eta=\frac{m^2 M_P^2 }{V_0 (1+\tilde\phi^2)},
\eeq
where we have introduced the rescaled scalar field $\tilde\phi$, which is dimensionless and defined so that $V=V_0(1+\tilde\phi^2)$.
Note that there are two limiting regimes: if $\tilde\phi\gg 1$,  the constant term is negligible and one recovers a power-law potential with $p=2$; if $\tilde\phi\ll 1$, $V_0$ dominates and the potential is extremely flat with $\epsilon\ll \eta$.

\subsection{The inflationary ``zoology''}
\subsubsection{Historical perspective}
The first model of inflation is usually traced back
to Alan Guth \cite{Guth:1980zm} in 
1981, although  one can see  the model  of 
Alexei Starobinsky \cite{Starobinsky:te} as a precursor.
 Guth's model, which is named today {\it old inflation}
 \index{old inflation}
is based on a first-order phase transition, from a false vacuum with 
non zero energy, which generates  an exponential inflationary phase, into 
a true vacuum with zero energy density. The true vacuum phase appears 
in the shape of bubbles via quantum tunneling. 
The problem with this inflationary model is that, in order to get sufficient
inflation to solve the problems of the standard model mentioned earlier, 
the nucleation rate must be sufficiently small; but, then, the bubbles never 
coalesce because the space that separates the bubbles undergoes 
inflation and expands too rapidly.  Therefore, the first model of inflation 
is not phenomenologically viable.

After this first and unsuccessful attempt, a new generation of inflationary 
models appeared, usually denoted {\it new inflation} models\index{new inflation}
\cite{new-inflation}. They rely 
 on a second order phase transition, based on thermal corrections of the 
effective potential and thus assume that the scalar field is in 
thermal equilibrium.

This hypothesis of thermal equilibrium 
was given up in the third generation of models, initiated 
by Andrei Linde, and whose generic name is \index{chaotic inflation}
{\it chaotic inflation} \cite{Linde:gd}.
This allows to use extremely simple potentials, quadratic or quartic, which 
lead to inflationary phases when the scalar field is displaced from the 
origin with values of the order of several Planck masses. 

In  the last few years, there has been an intense activity  in building 
inflationary models  based on high energy theories, in particular in the 
context of supersymmetry and string theory. Details can be found in several recent reviews~\cite{McAllister:2007bg,Burgess:2007pz,Kallosh:2007ig,Cline:2006hu,HenryTye:2006uv}.

\subsubsection{Classification}

There exist  a huge number of models of inflation.
 As far as {\it single-field} models 
are concerned\footnote{or at least {\it effectively} single field during inflation
(the hybrid models require a second field to {\it end} inflation as 
discussed earlier).}, it  is convenient to regroup them
into three broad categories:
\begin{itemize}
\item Large field models ($0<\eta\leq \epsilon$)
\index{large-field models}

The scalar field is displaced from its stable minimum by $\Delta \phi\sim M_P$.
This includes the  chaotic  models with monomial potentials
\beq
V(\phi)=\Lambda^4\left({\phi\over \mu}\right)^p,
\eeq
or the exponential potential
\beq
V(\phi)=\Lambda^4 \exp\left(\phi/ \mu\right),
\eeq
which have already been discussed.
\\

\item Small field models ($\eta<0<\epsilon $)
\index{small-field models}

In this type of models, 
the scalar field is rolling away from an unstable maximum of the potential.
This is a characteristic feature  of spontaneous symmetry breaking.
A typical potential is 
\beq
V(\phi)=\Lambda^4 \left[1-\left({\phi\over\mu}\right)^p\right],
\eeq
which can be interpreted as the lowest-order term in a Taylor expansion 
about the origin. Historically, this potential shape appeared in 
the so-called `new inflation' scenario.

A particular feature of these  models is that tensor modes 
are much more  suppressed with respect to scalar modes than in the 
large-field
models, as  it will be shown later.
\\

%%%%%%%%%%%%%%%%%%%%%%%%%%%%%%%%%%%%%%%%%%%%%%%%%%%%
\begin{figure}
\begin{center}
\includegraphics[width=4.8in]{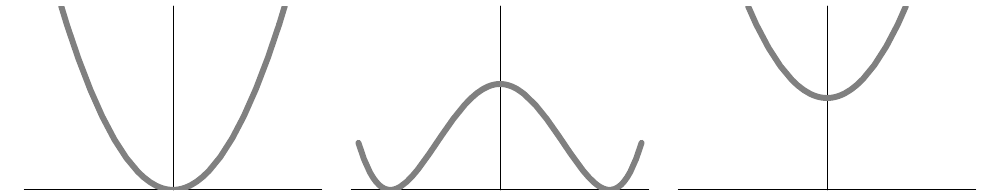}
\end{center}
\caption{Schematic potential for the three main categories of inflationary models: large-field models, 
small-field models, hybrid models.}
\end{figure}
%%%%%%%%%%%%%%%%%%%%%%%%%%%%%%%%%%%%%%%%%%%%%%%%%%%%

\item Hybrid models ($0<\epsilon<\eta$)

%%%%%%%%%%%%%%%%%%%%%%%%%%%%%%%%%%%%%%%%%%%%%%%%%%%%
%\begin{figure}
%\begin{center}
%\includegraphics[width=4.8in]{hybrid.eps}
%\end{center}
%\caption{Typical potential $V(\phi, \psi)$ for hybrid inflation.}
%\end{figure}
%%%%%%%%%%%%%%%%%%%%%%%%%%%%%%%%%%%%%%%%%%%%%%%%%%%%

Although a second scalar field is needed to end inflation, hybrid models
correspond effectively to single-field models with a 
potential 
characterized by $V''(\phi)>0$ and $0<\epsilon<\eta$.
A typical potential is
\beq
 V(\phi)=\Lambda^4 \left[1+\left({\phi\over\mu}\right)^p\right].
\eeq 
Once more, this potential can be seen as the lowest order in a Taylor 
expansion about the origin. 

In the case of hybrid models, the 
value $\phi_N$ of the scalar field as a function of the number of 
e-folds before the end of inflation is not determined by the above potential
and, therefore, $(\phi_N/\mu)$ can be considered as a freely adjustable 
parameter.

\end{itemize}

%%%%%%%%%%%%%%%%%%%%%%%%%%%%%%%%%%%%%%%%%%%%%%%%%%%%%%%%%%%%%%%%%%%%%%%%%%%
%%%%%%%%%%%%%%%%%%%%%%%%%%%%%%%%%%%%%%%%%%%%%%%%%%%%%%%%%%%%%%%%%%%%%%%%%%%
\section{Quantum fluctuations and ``birth'' of cosmological perturbations}
%%%%%%%%%%%%%%%%%%%%%%%%%%%%%%%%%%%%%%%%%%%%%%%%%%%%%%%%%%%%%%%%%%%%%%%%%%%
%%%%%%%%%%%%%%%%%%%%%%%%%%%%%%%%%%%%%%%%%%%%%%%%%%%%%%%%%%%%%%%%%%%%%%%%%%%
\index{quantum fluctuations}

So far, we have concentrated our attention on strictly homogeneous and 
isotropic aspects of cosmology. Of course, this idealized version, 
although extremely useful, is not sufficient to account for real 
cosmology and it is now time to turn to the study of deviations from 
homogeneity and isotropy. 

In cosmology, inhomogeneities grow because of the attractive nature of 
gravity, which implies that inhomogeneities were much smaller in the past.
As a consequence, for most of their evolution, inhomogeneities can 
be treated as {\it linear perturbations}. The linear
treatment  ceases to be valid  on small 
scales in our recent past, hence the difficulty to reconstruct the 
primordial inhomogeneities from large-scale structure, but it is 
quite adequate to describe  the fluctuations of the CMB at the time 
of last scattering. This is  the reason why the CMB is currently 
the best observational probe of primordial inhomogeneities.

In this section, we concentrate on the perturbations of the inflaton and show how 
the accelerated expansion during inflation converts its {\it initial 
vacuum quantum fluctuations} into ``macroscopic'' cosmological perturbations
(see \cite{mc,gp,staro,hawking,Mukhanov:1985rz,Mukhanov:1988jd} for some of the historical works). In this sense, inflation 
 provides us with ``natural'' initial conditions. We will also see how the perturbations of the inflaton can be translated into 
 perturbations of the geometry. 

\subsection{Massless scalar field in  de Sitter}
\index{de Sitter}
As a warming-up, it is instructive to  discuss
 the case of a massless scalar field
in a  de Sitter universe, described by a  cosmological  metric  with exponential 
expansion,
\beq
ds^2=-dt^2+a^2(t)d\x^2, \quad a(t)=e^{Ht}.
\eeq
It turns out it is more convenient to use, instead of the cosmic time $t$, 
a conformal time $\tau$, defined by 
\beq
\tau=\int \frac{dt}{a(t)},
\eeq
so that the metric takes the particularly simple form
\beq
\label{conformal_metric}
ds^2=a^2(\tau)\left[-d\tau^2+d\x^2\right].
\eeq
In the de Sitter case, the conformal time is given by
\beq
\tau=-\frac{e^{-Ht}}{H}=-\frac{1}{aH},
\eeq
so that the scale 
factor in terms of $\tau$ is simply
\beq
a(\tau)=-{1\over H\tau}.
\eeq
The conformal time is here negative (so that the scale factor is positive) and 
goes from $-\infty$ to $0$.

The action for a massless scalar field   is given by 
\beq
\label{action_dS}
S=\int d^4x \sqrt{-g}\left(-{1\over 2}\partial_\mu\phi \partial^\mu\phi\right)
 =\int d\tau \, \,  d^3x\, \, 
 a^4\left[{1\over 2 a^2}{\phi'}^2-{1\over 2 a^2}{\vec\nabla \phi}^2\right],
\eeq
where we have substituted in the action the cosmological metric  
(\ref{conformal_metric}) and where a prime denotes a derivative with respect to the conformal time $\tau$. 
Note that, whereas we still allow for spatial
variations of the scalar field, i.e. inhomogeneities, we will assume here, 
somewhat  inconsistently, that the geometry is completely fixed as 
homogeneous. We will deal later with the question of  the metric perturbations.

 It is possible to eliminate the factor $a^2$ in front of the kinetic term $\phi'^2$ by introducing the new function
\beq
u=a\, \phi.
\eeq
This will generate a term proportional to $u u'$  but one can get rid of  it by an integration by parts. The action (\ref{action_dS}) 
 can then be rewritten as  
\beq
S={1\over 2}\int d\tau \, \, d^3x \,  
\left[{u'}^2-  {\vec\nabla u}^2 +{a''\over a}u^2\right].
\eeq
The first two terms  are familiar since they are the same as 
in 
the action for a free massless scalar field in Minkowski spacetime.
The fact that our scalar field here lives in de Sitter spacetime rather 
than Minkowski has been reexpressed as a {\it time-dependent 
effective mass} 
\beq
m^2_{eff}=-{a''\over a}=-{2\over \tau^2}.
\eeq

Let us now proceed  to the quantization of  the scalar field $u$ by using 
the standard procedure of quantum field theory.
One first turns $u$ into a quantum field denoted $\hat u$, which we 
expand in Fourier space as 
\index{quantization}
\beq
\label{Fourier_quantum}
\hat u (\tau, \vec x)={1\over (2\pi)^{3/2}}\int d^3k \left\{{\hat a}_{\vec k} u_k(\tau) e^{i \vec k.\vec x}
+ {\hat a}_{\vec k}^\dagger u_k^*(\tau) e^{-i \vec k.\vec x} \right\},
\eeq
where  the $\hat a^\dagger$ and  $\hat a$ are 
 creation and annihilation operators, satisfying the 
usual commutation rules 
\beq
\label{a}
\left[ {\hat a}_{\vec k}, {\hat a}_{\vec k'}\right]= 
\left[ {\hat a^\dagger}_{\vec k}, {\hat a^\dagger}_{\vec k'}\right]= 0, 
\quad
\left[ {\hat a}_{\vec k}, {\hat a^\dagger}_{\vec k'}\right]= \delta(\vec k-\vec k').
\eeq
The function $u_k(\tau)$ is a complex time-dependent function that  
must satisfy the {\it classical} equation of motion in Fourier space, namely
\beq
\label{eom_u}
u_k''+\left(k^2-{a''\over a}\right)u_k=0, 
\eeq
which is simply the equation of motion for an oscillator with 
a time-dependent mass. In the case of a massless scalar field in 
Minkowski spacetime, this effective mass is zero ($a''/a=0$) 
and one usually takes 
\beq
u_k=\sqrt{\frac{\hbar}{ 2k}}e^{-ik\tau} \quad ({\rm Minkowski}),
\eeq
 where the choice for the 
normalization factor will be clear below.
In the case of de Sitter, one can  solve explicitly the above 
equation with $a''/a=2/\tau^2$ and the general solution is given by
\beq
\label{general_solution}
u_k=\alpha e^{-ik\tau }\left(1-{i\over k\tau}\right)
+\beta e^{ik\tau }\left(1+{i\over k\tau}\right).
\eeq

Canonical quantization  consists in imposing the following commutation rules
on the $\tau=$constant hypersurfaces:
\beq
\left[\hat u(\tau, \vec x), \hat u(\tau, \vec x')\right]=
 \left[\hat\pi_u(\tau, \vec x), \hat\pi_u(\tau, \vec x')\right]=0
\eeq
and 
\beq
\label{canonical}
\left[\hat u(\tau, \vec x), \hat\pi_u(\tau, \vec x')\right]=i\hbar \delta(\vec x-\vec x'), 
\eeq
where $\pi_u\equiv \delta S/\delta u'$ is the conjugate momentum of $u$.
In the present case, 
 $\pi_u=u'$ since the kinetic term is canonical.

Subtituting the expansion (\ref{Fourier_quantum}) in 
the commutator (\ref{canonical}), and using the commutation rules 
for the creation and annihilation operators (\ref{a}), 
one obtains the relation
\beq
u_k {u'_k}^*-u_k^*u'_k=i\hbar,
\label{wronskien}
\eeq
which determines the normalization of the Wronskian.

The choice of a specific  function $u_k(\tau)$  
corresponds to a particular 
 prescription for the physical vacuum $| 0 \rangle$, 
defined by
\beq
{\hat a}_{\vec k}|0\rangle=0.
\eeq 
A different choice for $u_k(\tau)$ is associated to a different decomposition
into creation and annihilitation modes and thus to a different vacuum. 

Let us now  note that the wavelength associated with 
a given mode $k$ can always be found {\it within} the Hubble radius 
provided one goes sufficiently far  backwards  in time, since 
the comoving Hubble radius is shrinking during inflation. 
In other words, for $|\tau|$ sufficiently big,
one gets $k|\tau|\gg 1$. 
Moreover, for a wavelength smaller than the Hubble radius, 
one can neglect the influence of the curvature of spacetime and the 
mode behaves as in a Minkowski spacetime, as can also be checked explicitly 
with the equation of motion (\ref{eom_u}) 
(the effective mass is negligible for 
$k|\tau|\gg 1$).
Therefore, the most natural physical prescription is to take  the particular
solution that corresponds to the usual Minkowski vacuum, i.e. 
$u_k\sim \exp(-ik\tau)$, in the limit $k|\tau|\gg 1$.
In view of (\ref{general_solution}), this corresponds to the choice
\beq
u_k=\sqrt{\hbar\over 2k}e^{-ik\tau }\left(1-{i\over k\tau}\right), 
\label{u_k}
\eeq 
where the coefficient has been determined by the normalisation condition
(\ref{wronskien}). 
This choice, in the jargon of quantum field theory on curved 
spacetimes, corresponds to the {\it Bunch-Davies vacuum}.

Finally, one can compute the {\it correlation function} for the  
scalar field $\phi$ in the vacuum state defined above. When Fourier 
transformed, the correlation function defines the {\it power 
spectrum} $\P_\phi(k)$:
\beq
\langle 0| \hat \phi(\vec x_1) \hat\phi(\vec x_2)|0\rangle= 
\int d^3 k \, \, e^{i \vec k.(\vec x_1-\vec x_2)}
{\P_\phi(k)\over 4\pi k^3}.
\eeq
\index{power spectrum}
Note that the homogeneity and isotropy of the quantum field 
is used implicitly in the definition of the power spectrum, which is 
``diagonal'' in Fourier space (homogeneity) and depends only on the 
norm of  $\vec k$ (isotropy).
In our case, we find 
\beq
\label{ps_phi}
2\pi^2k^{-3}\P_\phi= {|u_k|^2\over a^2},
\eeq
which gives in the limit $k|\tau|\ll 1$, i.e. when the wavelength is {\it larger than  the 
Hubble radius}, 
\beq
\P_\phi(k)\simeq \hbar \left({H\over 2\pi}\right)^2 
\qquad (k\ll aH)\, .
\eeq
Note that, in  the opposite limit $k|\tau|\gg 1$, 
one recovers the usual vacuum fluctuations in 
Minkowski, with $\P_\phi(k)=\hbar (k/2\pi a)^2$.

We have used  a quantum description of the scalar field. 
But  cosmological 
perturbations in the standard cosmological eras are usually  described  by {\it classical random 
fields}. 
Roughly speaking, the transition between the quantum and classical (although
stochastic) descriptions makes sense  when the perturbations  exit 
 the Hubble radius.  Indeed
 each of the terms in the 
 Wronskian (\ref{wronskien}) is roughly of the order 
 $\hbar/2(k\tau)^3$ in the super-Hubble limit and the non-commutativity 
can then be neglected. In this sense, one can see the exit outside   the 
Hubble radius as a quantum-classical transition, although some refinement 
is  required to make this statement more precise (see e.g. \cite{Polarski:1995jg}).

%%%%%%%%%%%%%%%%%%%%%%%%%%%%%%%%%%%%%%%%%%%%%%%%%%%%%%%%%%%%%
\subsection{Quantum fluctuations with metric perturbations}
%%%%%%%%%%%%%%%%%%%%%%%%%%%%%%%%%%%%%%%%%%%%%%%%%%%%%%%%%%%%%
Let us now move to the more realistic case of a perturbed inflaton field  
living in a {\it perturbed} cosmological geometry. 
In fact,  Einstein's equations imply 
that scalar field fluctuations must  necessarily coexist  with {\it metric 
fluctuations}. 
A correct treatment, either classical or quantum, must therefore involve 
both the scalar field perturbations and metric perturbations. We  thus need to resort 
to the theory of relativistic cosmological perturbations, which we briefly present below (more details can be found in e.g.
\cite{Bardeen:1980kt,ks,mfb,cargese,Malik:2008im}).

\subsubsection{Linear perturbations of the metric}
\index{metric perturbations}
The most general
linear perturbation about the homogenous metric can be expressed as
\beq
ds^2=a^2\left\{ -(1+2A)d\tau^2+2B_idx^id\tau+
\left(\d_{ij}+h_{ij}\right) dx^idx^j \right\} 
\label{metpert},
\eeq
where we have assumed, for simplicity,  a  spatially flat  background metric\footnote{This is all the more justified given that the metric in the early Universe was closer to a spatially flat metric than our present metric, which is itself indistiguishible from a flat geometry, according to observations.}.
We have introduced a time plus space decomposition of the perturbations.
The indices $i$, $j$ stand for {\it spatial} indices and the perturbed 
quantities   defined in (\ref{metpert}) 
can  be seen as three-dimensional tensors, for which the 
indices can be  lowered (or raised) by the spatial metric 
$\d_{ij}$ (or its inverse).

It is very convenient to separate the perturbations into three categories, 
the so-called ``scalar'', ``vector'' and ``tensor'' modes.
For example, a spatial vector field  $B^i$ can be decomposed uniquely into
a longitudinal part and a transverse part, 
\beq
B_i=\partial_i B+\B_i, \quad \partial_i\B^i=0,
\label{Bi}
\eeq
where the longitudinal part is curl-free and can thus be expressed as 
a gradient, and the transverse part is divergenceless. This yields
one ``scalar'' mode, $B$ , and two ``vector'' modes  $\B^i$ (the index $i$ 
takes three values but the divergenceless condition implies that only 
two components are independent).

A similar procedure applies to the symmetric tensor $h_{ij}$, which can 
be decomposed as 
\beq
h_{ij}=2C\, \d_{ij}+2 \partial_i\partial_j E+2\partial_{(i}E_{j)}
+\overline{E}_{ij}, 
\label{hij}
\eeq
with $\overline{E}^{ij}$ transverse and traceless (TT), 
i.e. $\partial_i\overline{E}^{ij}=0$ (transverse) and
$\overline{E}^{ij}\d_{ij}=0$ (traceless), and $E_i$ transverse.
The parentheses around the indices denote symmetrization, namely
$2\partial_{(i}E_{j)}\equiv \partial_{i}E_{j} + \partial_{j}E_{i}$.
We have thus defined 
 two scalar modes, $C$ and $E$, two vector modes, $E_i$, 
and two tensor 
modes,  $\E_{ij}$.

\subsubsection{Coordinate transformations}
\index{coordinate transformations}
The metric perturbations, introduced  in  (\ref{metpert}), are 
modified in a coordinate transformation  of the form
\beq x^\alpha\rightarrow x^\alpha+\xi^\alpha, \qquad \xi^\alpha=(\xi^0,\xi^i) 
\label{transjauge}.
\eeq
It can be shown that the change of the metric components can be expressed 
as 
\beq
\delta g_{\mu\nu}\rightarrow \delta g_{\mu\nu}
-2\nabla_{(\mu}\xi_{\nu)}\, ,
\eeq
using the symbol $\nabla$ for the four-dimensional covariant derivative, where the variation due the coordinate transformation is defined  for the {\it same} old and new coordinates (and thus at different physical points).

The above variation can be decomposed into 
individual variations  for the various components of the metric defined 
earlier. One finds
\begin{eqnarray}
A&\rightarrow&A-{\xi^0}'-\h\xi^0 \label{gt1}\\
       B_i&\rightarrow&B_i+\partial_i\xi^0-\xi_i' \label{gt2}\\
          h_{ij}&\rightarrow&h_{ij}- 2 \left(\partial_{(i}\xi_{j)}
-\h\xi^0\d_{ij}\right)\, , \label{gt3}
\end{eqnarray}
where $\h\equiv a'/a$.

The effect of a coordinate transformation can also be decomposed along
the scalar, vector and tensor sectors introduced earlier.
The generator $\xi^\alpha$ of the coordinate transformation can indeed be 
written as 
\beq  
\xi^\alpha=(\xi^0,\partial^i\xi+\overline{\xi}^i), 
\eeq
with $\overline{\xi}^i$ transverse, 
which  shows explicitly that $\xi^\alpha$ contains two scalar 
components,  $\xi^0$ and $\xi$, and two vector components,
$\overline{\xi}^i$. The transformations (\ref{gt2}-\ref{gt3}) are then 
decomposed into :
\begin{eqnarray}
 B&\rightarrow &B+\xi^0-\xi' \nonumber\\
C&\rightarrow &C-\h\xi^0  \nonumber\\
 E&\rightarrow &E-\xi \label{transfjauge} \\
\overline{B}^i&\rightarrow &\overline{B}^i-{{\overline{\xi}}^i}' \nonumber\\
 E^i&\rightarrow &E^i-\overline{\xi}^i. \nonumber
 \end{eqnarray}
The tensor perturbations remain unchanged since $\xi^\alpha$ does not
contain any tensor component.

To summarize, the whole system scalar field plus gravitation is described by eleven perturbations. They can be decomposed into 
five scalar quantities: $A$, $B$, $C$ and $E$ from the metric and $\delta\phi$; four vector quantities $\bar{B}^i$ and $\bar{E}^i$;
two tensor quantities: the two polarizations of $E^{TT}_{ij}$. However, these quantities are physically redundant since the same physical situation can be described by different sets of values of these perturbations, provided they are related by the coordinate transformations described above.  

One would thus like to identify the {\it true} degrees of freedom, i.e. the physically independent quantities characterizing the system. One can reduce the effective number of degrees of freedom by using  the four coordinate transformations, which consist of  two scalar transformations  and two vector transformations as we saw earlier. 
Moreover, Einstein's equations contain nondynamical equations, i.e. constraints, which are also the consequence of the invariance by coordinate transformations. They can be decomposed into two scalar constraints and two vector constraints.
 The situation for the scalar, vector and tensor sectors, respectively,  is summarized in Table~\ref{table}. By taking into account the coordinate changes and the constraints, one finds  three true degrees of freedom: two polarizations of the gravitational waves and one scalar degree of freedom. If matter was composed of $N$ scalar fields, one would get $N$ scalar degrees of freedom in addition to the two tensor modes.

\begin{table}[t]
\begin{center}
\begin{tabular}{|c|c|c|c|c|c|}
\hline
 & Metric & Scalar field & Gauge choice & Constraints & True d.o.f.
\\ \hline
S & 4 & 1 & -2 & -2 & 1
\\ 
V & 4 & 0 & -2 & -2 & 0
\\ 
T & 2 & 0 & \ 0 & \ 0 & 2
\\ \hline
\end{tabular}
\caption{Counting of the degrees of freedom in the scalar, vector and tensor sectors.}
\label{table}
\end{center}
\end{table}

In a coordinate transformation, the scalar field perturbation is also modified, according to 
\beq
\delta\phi \rightarrow \delta\phi-\phi'\xi^0\,.
\eeq
In single-field inflation, there are thus two natural choices  of gauge to describe the scalar perturbation.  
The first is to work with hypersurfaces that are flat, i.e. $C=0$, in which case we will denote the scalar field perturbation by $Q$, i.e. 
\beq
Q=\delta\phi_{_{C=0}}.
\eeq
The other choice is to work with hypersurfaces where the scalar field is uniform, i.e. $\delta\phi=0$, in which case the scalar degree of freedom is embodied by the metric perturbation $C_{\delta\phi=0}$. In other words, the true scalar degree of freedom can be represented either as a pure matter perturbation or a pure metric perturbation. In the general case, we have
\beq
Q=\delta\phi-\frac{\phi'}{\h}C\, ,
\eeq
which is a gauge-invariant combination (often called the Mukhanov-Sasaki variable~\cite{Mukhanov:1985rz,Sasaki:1986hm}).

\subsection{Quantizing the scalar degree of freedom}
\def\p{\phi}
\def\half{\frac12}

In order to quantize the true scalar degree of freedom, one needs the action that governs its dynamics. Let us first note that the {\it linearized} equations of motion for the coupled system \{gravity + scalar field\} are obtained from the expansion of the full action at {\it second-order} in the perturbations. Indeed the equations for the linear perturbations correspond to the Euler-Lagrange equations derived from a quadratic Lagrangian. 
In our case, the difficulty is that there are several scalar perturbations that are not independent. In order to quantize this coupled system, one can work  directly with the second-order Lagrangian \cite{mfb}, or resort to a Hamiltonian approach \cite{l94,Anderegg:1994xq}. 

The modern approach, introduced by Maldacena~\cite{Maldacena:2002vr} to study perturbations beyond  linear order,  is based on the  Arnowitt-Deser-Misner (ADM) \index{ADM formalism} formalism~\cite{ADM}. 
In the ADM approach,  the metric is written in the form
\beq
ds^2=-N^2 dt^2 +h_{ij} (dx^i+N^i dt)(dx^j+N^j dt)\,
\label{metric_ADM}
\eeq
where $N$  is called the lapse function and $N^i$ the shift vector.
The full action for the scalar field and  gravity
\beq
\label{action_scalar_field_Einstein}
S=\int d^4x\sqrt{-g}\left(-{1\over 2}\partial^\mu\phi\partial_\mu\phi
-V(\phi)\right) +\frac{M_P^2}{2}\int d^4x\sqrt{-g} R
\eeq
 becomes, after substitution of (\ref{metric_ADM}), 
\beq
S=\int dt d^3x \sqrt{h}N\, \left[\frac{\V^2}{2N^2} -\frac{1}{2}h^{ij} \partial_i \phi \partial_j \phi-V(\phi)\right]+ \frac{M_P^2}{2} \int dt d^3 x \frac{\sqrt{h}}{N}(E_{ij} E^{ij}-E^2)\,,
\label{action-ADM}
\eeq
where $h=$ det$(h_{ij})$,
\beq
\V\equiv \dot \phi -N^j \partial_j \phi\,.
\label{speed}
\eeq
 and  the symmetric tensor $E_{ij}$, defined by 
\beq
E_{ij}\equiv \frac12 \dot{h}_{ij}-N_{(i|j)}\,,
\label{Eij}
\eeq
(the symbol $|$ denotes the spatial covariant derivative associated with 
the spatial metric $h_{ij}$)
is proportional to the extrinsic curvature of the spatial slices.

The variation of the action with respect to $N$ yields
the energy constraint,
\index{energy constraint}
\beq
\frac{\V^2}{2N^2}+\frac{1}{2}h^{ij} \partial_i \phi \partial_j \phi+V(\phi)
+\frac{M_P^2}{2 N^2}(E_{ij} E^{ij}-E^2)=0\, ,
\label{N-constraint}
\eeq
while the variation of the action with respect to the shift $N^i$ gives 
the momentum constraint, 
\index{momentum constraint}
\beq
 M_P^2 \left( \frac{1}{N} (E^j_i - E \delta^j_i) \right)_{|j} =
  \frac{\V}{N}  \partial_i \phi .
  \label{Ni-constraint}
\eeq

In order to study the linear perturbations about the FLRW background, 
we now restrict ourselves to  the flat gauge, which corresponds to the choice
\beq
h_{ij}=a^2(t) \delta_{ij}.
\eeq 
The scalar fields on the corresponding flat hypersurfaces can be decomposed as
\beq
\phi = \bar\phi + Q
\eeq
 where  $\bar\phi$ is the spatially homogeneous background value of the scalar field and  $Q$ represents 
 its perturbation (on flat hypersurfaces). In the following, we will often omit the bar and  simply write the homogeneous value as $\phi$, unless this generates  ambiguities.

We can also write the (scalarly) perturbed lapse and shift as
\beq
N = 1 + \alpha, \qquad  N_i = \beta_{,i}\, ,
\eeq
where the linear perturbations $\alpha$ and $\beta$ are determined in terms of the scalar field perturbation 
$Q$ by solving the linearized constraints.
At first-order,  the momentum constraint implies
\beq
  \label{alpha}
  \alpha = \frac{\dot\phi}{2M_P^2\, H}\, Q\, ,
\eeq
while the energy constraint gives $\partial^{2}\beta$ in terms of $Q$ and $\dot Q$.

\subsection{Second order action}
We now expand the action, up to quadratic order,  in terms of the linear perturbations. This action can be written solely in terms of the physical degree of freedom $Q$ by substituting the expression (\ref{alpha}) for $\alpha$  (it turns out that $\beta$ disappears of the second order action, after an integration by parts).   
The second order action can be  written in the rather simple form
\begin{eqnarray}
S_{(2)}&=& \half \int dt \,\dn{3}{x}\,    a^3 \left[ {\dot Q}^2-\frac{1}{a^2}\partial_i Q \partial^i Q
  - {\cal M}^2 Q^2 \right]\,,
\end{eqnarray}
with the effective (squared) mass
\begin{eqnarray}
{\cal  M}^2  &=& V''-\frac{1}{a^3}\frac{d}{dt}\left(\frac{a^3}{H}\dot\phi^2\right)\, .
\end{eqnarray}

As we did earlier, it is convenient to use the conformal time $\tau$ and to introduce the canonical degree of freedom
\beq 
\label{v}
v=a\,  Q
\eeq
which leads to the action
\beq
S_v={1\over 2}\int d\tau \, \, d^3x\, \left[{v'}^2+\partial_i v \partial ^i v
+{z''\over z}v^2\right],
\eeq
with
\beq
\label{z}
z=a\, {\phi'\over \h}.
\eeq
This action is 
analogous to that of a scalar field in Minkowski spacetime 
with a time-dependent mass. The  situation is quite similar to what we obtained
 previously, with the notable difference that the 
effective time-dependent mass is now
$z''/z$,  instead of $a''/a$.  

The quantity we will be eventually interested in is the comoving 
curvature perturbation $\R$, which is related to the canonical variable 
$v$ by the relation 
\beq
v=z\, \R.
\eeq
Since, by analogy with (\ref{ps_phi}),
 the power spectrum for $v$ is given by
\beq
2\pi^2 k^{-3}{\cal P}_v(k)= |v_k|^2, 
\eeq
the corresponding power spectrum for $\R$ is found to be 
\beq
\label{ps_R}
2\pi^2 k^{-3}{\cal P}_\R(k)= {|v_k|^2\over z^2}.
\eeq

In the case of an inflationary phase in the {\it slow-roll} approximation, 
 the evolution of 
 $\phi$ and of  $H$ is much slower than 
 that of the scale factor $a$. 
Consequently, one gets approximately 
\beq
{z''\over z}\simeq {a''\over a},  \qquad{\rm (slow-roll)}
\eeq
and all results  obtained previously for  $u$
apply directly  to our variable $v$ in the slow-roll approximation.
This implies that the properly normalized function corresponding to 
the Bunch-Davies vacuum is approximately given by 
\beq
v_k\simeq \sqrt{\hbar\over 2k}e^{-ik\tau }\left(1-{i\over k\tau}\right). 
\label{v_k}
\eeq 
In the super-Hubble limit $k|\tau| \ll 1$ the function $v_k$ behaves 
like
\beq
v_k\simeq - \sqrt{\hbar\over 2k}\, {i\over k\tau} \simeq 
i\sqrt{\hbar\over 2k}\, {aH\over k}, 
\eeq 
where we have used $a\simeq - 1/(H\tau)$.

Consequently,  combining (\ref{ps_R}), (\ref{z}) and (\ref{v_k}) and reintroducing the cosmic time gives the power spectrum for $\R$, on scales larger than the Hubble radius,
\beq
\P_\R\simeq {\hbar\over 4\pi^2}\left({H^4\over \dot\phi^2}\right)_{k=aH}=\frac{\hbar}{2\mP^2\epsilon_*}\left(\frac{H_*}{2\pi }\right)^2 
\label{power_S}
\eeq
where we have used $\epsilon\equiv -\dot H/H^2$ in the second equality, and the subscript $*$ means that the quantity is evaluated at Hubble crossing ($k=aH$). 
This is the main result for the spectrum 
of  scalar cosmological perturbations 
generated from vacuum fluctuations during a slow-roll inflation phase.

%%%%%%%%%%%%%%%%%%%%%%%%%%%%%%%%%%
\subsection{Gravitational waves}
%%%%%%%%%%%%%%%%%%%%%%%%%%%%%%%%%%
\index{gravitational waves}
\index{tensor modes}
We have focused so far  our attention on scalar perturbations, which 
are the most important in cosmology. Tensor perturbations, or primordial 
gravitational waves, if ever detected in the future, would be a remarkable 
additional probe of the early universe. In the inflationary scenario, 
like scalar perturbations, primordial 
gravitational waves are generated from vacuum quantum fluctuations~\cite{Starobinsky:1979ty}.
 Let us now explain briefly how.

The action expanded at  second order in the perturbations contains a 
tensor part, 
given by 
\beq
S^{(2)}_g=\frac{M_P^2}{8}\int  d\tau\,  d^3x \, a^2 
\eta^{\mu\nu}\partial_\mu \bar E^i_j\partial_\nu\bar E^j_i,
\eeq
where $\eta^{\mu\nu}$ denotes the Minkoswki metric.
Apart from the tensorial nature of $E^i_j$, this action is quite 
similar to that of a scalar field in a FLRW universe (\ref{action_dS}), 
up to a renormalization 
factor $M_P/2$.
The decomposition
\beq
a\bar E^i_j=\sum_{\lambda=+,\times}
\int {d^3k\over (2\pi)^{3/2}} 
 v_{k,\lambda}(\tau)\epsilon^i_j({\vec k};\lambda) e^{i \vec k.\vec x}
\eeq
where the $\epsilon^i_j({\vec k};\lambda)$ are the polarization tensors,
shows that the gravitational waves are essentially equivalent to two 
massless
scalar fields (for each polarization) $\phi_\lambda=M_P\bar E_\lambda /2$.

The total power spectrum is thus immediately deduced from (\ref{ps_phi}) and reads
\beq
{\cal P}_T=2\times {4\over \mP^2}\times \hbar\left({H\over 2\pi}\right)^2, 
\eeq
where the first factor comes from the two polarizations, the 
second from the renormalization with respect to a canonical scalar field, 
the last term being the power spectrum for a scalar field derived earlier.
In summary, the tensor power spectrum is
\beq
{\cal P}_T=\frac{8\hbar}{\mP^2}\left(\frac{H_*}{2\pi}\right)^2,
\label{power_T}
\eeq
where the subscript $*$ recalls that the Hubble parameter, which can be slowly evolving
during inflation, must be evaluated when the relevant scale exited the 
Hubble radius during inflation. 

A measurement of the tensor amplitude (\ref{power_T}) gives direct access, in this context, to the energy scale $H_*$ during inflation, in contrast with the scalar amplitude (\ref{power_S}) which depends on the slow-roll parameter $\epsilon_*$ as well. The tensor to scalar ratio,
 \beq
\label{r}
r\equiv
{{\cal P}_T\over {\cal P}_\R}=16\, \epsilon_*\, ,
\eeq
 is proportional to the 
slow-roll parameter.

%%%%%%%%%%%%%%%%%%%%%%%%%%%%%%%%%%%%%%%%%%%%%%%%%%%%
%%%%%%%%%%%%%%%%%%%%%%%%%%%%%%%%%%%%%%%%%%%%%%%%%%%%
\section{From inflation to the standard era}
%%%%%%%%%%%%%%%%%%%%%%%%%%%%%%%%%%%%%%%%%%%%%%%%%%%%
%%%%%%%%%%%%%%%%%%%%%%%%%%%%%%%%%%%%%%%%%%%%%%%%%%%%
Once the perturbations have been computed during inflation, one must relate them to  perturbations in the standard radiation dominated era, where they will be used as ``initial conditions''. A priori, one could think that it is necessary to follow the details of how the inflaton is converted into ordinary particles in order to establish this relation. In fact, all these details turn out to be  irrelevant, at least in the case of single field inflation,  because there exists a conservation law for scales larger than the Hubble radius, which is the case for 
all relevant scales at the end of inflation.

\subsection{Covariant approach}
\index{covariant formalism}
Instead of  the traditional metric-based approach, we  use here a more geometrical approach to cosmological perturbations~\cite{Ellis:1989jt}, which will enable us to recover easily and intuitively the main useful results, not only for linear perturbations but also for non-linear perturbations.

Let us consider a spacetime with metric $g_{ab}$ and some perfect fluid characterized by its energy density $\rho$, its pressure $P$ and its four-velocity $u^a$. The corresponding energy momentum-tensor is given by
\beq 
T_{ab}=\rho u_a u_b+P(g_{ab}+u_au_b).
\eeq
Let us also introduce the expansion along the fluid worldlines,
\beq
 \Theta=\nabla_a u^a, 
\eeq
and the integrated expansion
\beq 
 \quad \alpha=\frac{1}{3}\int d\tau_p \,
\Theta \;,
\eeq
where $\tau_p$ is the proper time defined along the fluid worldlines.  In a FLRW spacetime, one would find $\Theta=3H$. Therefore, in the general case, one can interpret $\Theta/3$ as a local Hubble parameter and $S=\exp(\alpha)$ as a local scale factor, while $\alpha$ represents the local number of e-folds. 
 
As shown in \cite{Langlois:2005ii,Langlois:2005qp}, the conservation law for the energy-momentum tensor,
\index{energy-momentum conservation}
 \beq
\nabla_a T^{a}_{\ b}=0, 
 \eeq
implies that the covector
\beq
 \zeta_a\equiv
\nabla_a\alpha-\frac{\dot\alpha}{\dot\rho}\nabla_a\rho
\label{zeta_a}
\eeq
satisfies the relation
\beq
\label{dot_zeta}
\dot\zeta_a\equiv {\cal L}_u\zeta_a=
-\frac{\Theta}{3(\rho+p)}\left( \nabla_a p -
\frac{\dot p}{\dot \rho} \nabla_a\rho\right) \;,
\eeq
 where a dot denotes the time derivative defined as the  Lie derivative along $u^a$\footnote{For scalar quantities, this 
  is equivalent to an ordinary
derivative along $u^a$ (e.g. $\dot\rho\equiv u^a\nabla_a\rho$), but for $\zeta_a$, one has 
$\dot\zeta_a\equiv u^b\nabla_b \zeta_a+ \zeta_b\nabla_a u^b$.
}. This result is valid for any spacetime geometry and does not depend on Einstein's equations. 

The covector $\zeta_a$ can be defined for the global cosmological fluid or for any of the
individual cosmological fluids (the case of interacting fluids is discussed in \cite{Langlois:2006iq}). Using
the non-linear conservation equation
\beq
\dot\rho=-3\dot\alpha(\rho+P)\;,
\eeq
which follows from $u^b\nabla_a T^a_{\ b}=0$,
one can re-express  $\zeta_a$ in the form
\beq
\label{zeta_a2}
\zeta_a=\nabla_a\alpha+\frac{\nabla_a\rho}{3(\rho+P)} \;.
\eeq
If $w\equiv P/\rho$ is constant,
the above covector  is a total gradient and can be written as
\beq
\zeta_a=\nabla_a\left[\alpha+\frac{1}{3(1+w)}\ln \rho\right]\, .
\label{zeta_a_w}
\eeq

On scales larger than the Hubble radius, our definition agrees
with the non-linear curvature perturbation on uniform
density hypersurfaces as defined in
\cite{Lyth:2004gb} (see also \cite{Rigopoulos:2003ak})
 \beq \zeta = \delta N -
 \int_{\bar \rho}^{\rho} H \frac{d \tilde \rho}{\dot{ \tilde \rho}}
 =
 \delta N + \frac13\int_{\bar \rho}^{\rho} \frac{d
 \tilde{\rho}}{(1+w)\tilde{ \rho}}\;, 
 \eeq
where $N=\alpha$. The
above equation is simply the  integrated version of
(\ref{zeta_a}), or of (\ref{zeta_a2}).

%%%%%%%%%%%%%%%%%%%%%%%
\subsection{Linear conserved quantities}
%%%%%%%%%%%%%%%%%%%%%%%
Let us now introduce a coordinate system, in which the metric (with only scalar perturbations) 
reads
\beq
ds^2=a^2\left\{ -(1+2A)d\tau^2+2\partial_iB \, dx^id\tau+
\left[(1+2C)\d_{ij}+2\partial_i\partial_jE\right] dx^idx^j \right\} 
\label{metric_pert_scalar}.
\eeq
One can  decompose the fluid four-velocity as 
\beq
\label{u_pert}
u^\mu=\bar{u}^\mu+\delta u^\mu, \quad \delta u^\mu=\left\{-A/a, v^i/a\right\},\quad  v_i=\partial_iv+\bar{v}_i\, ,
\eeq
where $\bar{v}_i$ is transverse. 

At linear order, the spatial components of $\zeta_a$  are simply
\beq
\zeta_i^{(1)}=\partial_i\zeta^{(1)}, \qquad \zeta^{(1)}\equiv\delta\alpha-\frac{\bar{\alpha}'}{\bar{\rho}'}\delta\rho\, ,
\eeq
where a prime denotes a derivative with respect to $\tau$.
Linearizing (\ref{dot_zeta}) implies  that  the curvature perturbation \index{curvature perturbation}
on {\it uniform-energy-density hypersurfaces}, defined by 
\beq
\zeta=C-\h {\delta\rho\over  \rho'}=C+\frac{\delta\rho}{3(\rho+P)}\, 
\label{zeta}
\eeq
and originally introduced  in \cite{Bardeen:1983qw},
obeys the evolution equation (see also \cite{Wands:2000dp})
\beq
\zeta'=-{\h\over \rho + P}\d P_{nad}-{1\over 3}\nabla^2(E'+v),
\label{zetaprim}
\eeq
where $\d P_{nad}$ is the non-adiabatic part of the pressure perturbation, 
defined by
\beq
\d P_{nad}=\d P-c_s^2\d\rho.
\eeq
Note that $\zeta^{(1)}$ differs from $\zeta$ but they coincide when the spatial gradients can be neglected, for instance on large scales. 
The expression (\ref{zetaprim}) shows that $\zeta$ is conserved 
{\it on super-Hubble scales} in the case of {\it adiabatic} perturbations.

Another convenient quantity, 
which is sometimes used in the literature instead of 
$\zeta$, is the {\it curvature perturbation on comoving hypersurfaces}, 
which can be written in any gauge as 
\beq
\R=-C-{\h\over \rho+P} \delta q, \qquad \partial_i\delta q\equiv \delta_{(S)} T^0_i,
\label{R}
\eeq
where the subscript $(S)$ denotes the perturbations of scalar type. For a perfect fluid, $\delta q=(\rho+P) v$, where $v$ has been defined in (\ref{u_pert}).

One can relate the two quantities $\zeta$ and $\R$ by using the energy and momentum constraints, which were derived earlier in the ADM formalism. Linearizing (\ref{N-constraint}) and (\ref{Ni-constraint}) yields, respectively, 
\begin{eqnarray}
3 \h^2\delta N+a\h\partial^2\beta&=&-\frac{a^3}{2M_P^2}\delta \rho\\
\h\delta N&=&-\frac{a^3}{2M_P^2}\delta q\, .
\end{eqnarray}
Combining these two equations yields  the relativistic analog of the Poisson equation, namely
\beq
\label{poisson}
\partial^2\Psi=\frac{a^2}{2M_P^2}\left(\delta\rho-3\h \delta q\right)\equiv\frac{a^2}{2M_P^2}\delta\rho_c \, ,
\eeq
where we have replaced $\beta$ by the Bardeen potential $\Psi\equiv-C- \h(B-E')=-\h\beta$ and introduced 
the comoving energy density $\delta\rho_c\equiv\delta\rho-3\h\delta q$. 
Since 
\beq
\zeta=-\R+{\d\rho_c\over \rho+P}=-\R-{2\rho\over 3(\rho+P)}\left({k\over aH}\right)^2\Psi,
\eeq
one finds  that $\zeta$ and $\R$ coincide in the 
super-Hubble limit $k\ll aH$.

%%%%%%%%%%%%%%%%%%%%%%%%%%%%%%%%%%%%%%%%%%%%%%%%%%%%%%%%
\subsection{``Initial'' conditions for standard cosmology}
%%%%%%%%%%%%%%%%%%%%%%%%%%%%%%%%%%%%%%%%%%%%%%%%%%%%%%%%

In standard cosmology, the ``initial'' conditions for the perturbations are usually defined in the radiation dominated era around the time of nucleosynthesis, 
when  the main cosmological components are restricted to  the usual  photons, baryons,  neutrinos and  cold dark matter (CDM) particles. 
The scales  that are cosmologically relevant today correspond to lengthscales much larger than the Hubble radius at that time. 

Before the invention of inflation,  ``initial'' conditions were put ``by hand'', with the restriction that their late time consequences should be compatible with observations. Inflation now provides a precise prescription to determine these ``initial conditions''\footnote{although one must be aware that present cosmological scales can correspond to scales smaller than the Planck scale during inflation, suggesting the possibility of  trans-Planckian effects (see e.g.~\cite{Martin:2000xs}).}.  

Since several species are present, one needs to specify  the density perturbation of each species. 
A simplification arises in the case of {\it single field} inflation, since exactly the same cosmological history, i.e. inflation followed by the decay of the inflaton into the usual species,  occurs in all parts of our Universe, up to a small time shift depending on the perturbation of the inflaton in each region. 
As a consequence, even if the number densities of the various species vary from point to point, their ratio should be fixed, i.e.
\beq
\delta\left(n_A/n_B\right) =0\, ,
\label{nullratio}
\eeq
for any pair of  species denoted $A$ and $B$ (see e.g. \cite{Weinberg:2004kr} for a more detailed discussion). 
This is not necessarily true in {\it  multi-field inflation}, as the perturbations in the radiation era may depend on {\it different} combinations of the scalar field perturbations.

 The variation in the relative particle number densities between 
two species can be quantified by the quantity 
\beq
S_{A,B}\equiv {\d n_A\over n_A}-{\d n_B\over n_B}\, ,
\eeq
\index{entropy perturbation}
which is usually called the {\it entropy} perturbation between $A$ and $B$.
 When the equation of state for a given species is such that
$w\equiv P/\rho= const$,  one can reexpress the entropy
perturbation in terms of the density contrast, in the form
\beq
S_{A,B}\equiv {\d_A\over 1+w_A}-{\d_B\over 1+w_B}.
\eeq
It is convenient to choose a species of reference, for instance the 
photons, and to define the entropy perturbations of the other species 
relative to it. The quantities
\begin{eqnarray}
S_b&\equiv \d_b-{3\over 4} \d_\gamma, \\
S_c&\equiv \d_c-{3\over 4} \d_\gamma, \\
S_\nu&\equiv {3\over 4}\d_\nu-{3\over 4} \d_\gamma,
\end{eqnarray}
thus  define respectively the baryon, CDM and neutrino entropy perturbations.

For single field inflation, all these entropy perturbations vanish, $S_b=S_c=S_\nu=0$, and the primordial perturbations are said to be {\it adiabatic}. 
An adiabatic primordial perturbation is thus characterized by 
\index{adiabatic perturbation}
\beq
{1\over 4}\d_\gamma={1\over 4}\d_\nu={1\over 3}\d_b={1\over 3}
\d_c\, .
\eeq
Only one density constrast needs to be specified. However, since it is a gauge-dependent quantity, it is better to use the gauge-invariant quantity $\zeta$, i.e. the uniform density curvature perturbation, which is also equivalent to $-\R$, since we are on super-Hubble scales here. 
 
Note that the {\it adiabatic mode}, which is directly related to the curvature perturbation, is also called {\it curvature} mode. By contrast, the entropy perturbations can be non-zero even if the curvature is zero, and the corresponding modes are called {\it isocurvature} modes.

\subsection{Inflation and cosmological data}
Let us now discuss the confrontation of single-field inflation models with the current cosmological data. The main idea is that one can predict precisely the statistics of the CMB perturbations, once  the amplitude of the primordial perturbation as a function of  scale, $\R(k)$, is given, provided some choice for the cosmological parameters $\Omega_i$. In other words, the measurements of the CMB, together with other cosmological data, allow us to constrain both the cosmological parameters, which are numbers, and the primordial spectrum, which is a function (see e.g. \cite{durrer,Challinor:2009tp} for details on the CMB physics).
From the present data, one finds that the primordial spectrum is nearly  (although not quite) scale-invariant , with an amplitude
\beq
{\cal P}_\R^{1/2}\simeq 5\times 10^{-5}.
\eeq

In order to derive some constraints on the inflation models, it is useful to reexpress the scalar and tensor power spectra, respectively 
given in (\ref{power_S}) and (\ref{power_T}), in terms of the 
scalar field potential. This can be done by using the slow-roll 
equations 
(\ref{sr1}-\ref{sr2}). One finds for the scalar spectrum
\beq
\label{spectrum_obs}
{\cal P}_\R(k)={1\over 12\pi^2}\left({V^3\over \mP^6{V'}^2}\right)_{k=aH}\, ,
\eeq
the subscript meaning  that the term 
on the right hand side must be evaluated at {\it Hubble radius exit}
for the scale of interest.
The scalar spectrum can also be written in terms of the 
first slow-roll parameter defined in (\ref{epsilon}), 
in which case it reads
\beq
{\cal P}_\R(k)={1\over 24\pi^2}\left({V\over \mP^4\epsilon_V}\right)_{k=aH}\, .
\eeq
If $\epsilon_V$ is not much smaller than $1$, as in chaotic models, the observed amplitude (\ref{spectrum_obs}) implies that 
the typical energy scale during inflation is
\beq
V^{1/4}\sim 10^{-3}\mP\sim 10^{15}{\rm GeV}.
\eeq
As for the tensor power spectrum, it is given in terms of the scalar field potential by
\beq
{\cal P}_T={2\over 3\pi^2}\left({V\over \mP^4}\right)_{k=aH}.
\eeq

The scalar and tensor spectra are almost scale invariant but not quite
since the scalar field slowly evolves  during the inflationary phase.
In order to evaluate quantitatively this variation, it is convenient to
introduce a scalar  {\it spectral index} as well as a tensor one, 
defined respectively by  
\index{spectral index}
\beq
n_S(k)-1={d\ln {\cal P}_\R(k)\over d\ln k}, \quad 
n_T(k)={d\ln {\cal P}_T(k)\over d\ln k}.
\eeq
One can express the spectral indices in terms of the slow-roll 
parameters. For this purpose, let us note that, in the slow-roll 
approximation,  $d\ln k=d\ln(aH)\simeq d\ln a$, so that  
\beq
{d\phi\over d\ln a}=
{\dot\phi\over H}\simeq -{V'\over 3H^2}\simeq - M_P^2{V'\over V},
\eeq
where the slow-roll equations 
 (\ref{sr1}-\ref{sr2}) 
have been used. 
Therefore, one gets 
\beq
n_s(k)-1=2\eta_V- 6 \epsilon_V,
\eeq
where $\epsilon_V$ and $\eta_V$ are the two slow-roll parameters defined
in  (\ref{epsilon}) and (\ref{eta}). 
Similarly, one finds for the tensor spectral index
\beq
n_T(k)=-2\epsilon_V.
\eeq
Comparing with Eq.~(\ref{r}) yields the 
relation
\beq
r=-8 n_T,
\eeq
the  so-called {\it consistency
relation}
\index{consistency relation}
which relates  purely {\it observable}  quantities (at least in principle). This means 
that if one was able to observe the primordial gravitational 
waves and  measure the amplitude and spectral index of their
spectrum, a rather  formidable task, then one 
would be able to test directly the paradigm of single field slow-roll
inflation.

Let us finally go back to  the particular models which we have already considered, in order to establish the predictions of these models in the $(n_s, r)$ plane, where they can be easily  compared with the observational constraints. 
For the power-law potentials (\ref{pot_power-law}),  one finds, using (\ref{SR_param_power-law}),
\beq
n_s-1=-6\epsilon+2\eta=-2\frac{p+2}{p}\epsilon
\eeq
and 
\beq
r=16\epsilon=\frac{8p}{p+2}(1-n_s)\, .
\eeq
Moreover, 
\beq
\epsilon=\frac{p}{4N},
\eeq
where $N$ is the number of e-folds before the end of inflation when the scales of cosmological interest crossed out the Hubble radius. 
Therefore, the observational prediction for a model with a power-law potential lie on a line in the $(n_s,r)$ plane, the precise point depending on the number of e-folds when the perturbations were generated.

For an exponential potential (\ref{potential_exponential}), one finds, using (\ref{SR_param_exponential}),
\beq
n_s-1=-\frac{2}{q}, \qquad
r=\frac{16}{q}.
\eeq
The prediction in the $(n_s, r)$ plane thus depends only on the parameter in the exponential of the potential, but not on the number of e-folds as in the previous case.

For potentials (\ref{hybrid}) like in hybrid inflation, one finds 
\beq
\eta=\frac{1+\tilde\phi^2}{\tilde\phi^2}\epsilon,
\eeq
and 
\beq
r=8\frac{\tilde\phi^2}{2\tilde\phi^2-1}(1-n_s).
\eeq

One can proceed in a similar way for any model of inflation and thus be able to confront it with observational data. 
In general, it is worth noticing that a significant amount of gravitational waves, and thus a detectable $r$, requires a variation of the inflaton of the order of the Planck mass during inflation~\cite{Lyth:1996im}. 

%%%%%%%%%%%%%%%%%
%%%%%%%%%%%%%%%%
\section{More general inflationary scenarios}
%%%%%%%%%%%%%%%%%
%%%%%%%%%%%%%%%%%%
So far, the simplest models of inflation are compatible with the available data but it is instructive to study more refined models for at least two reasons. First, models inspired by high energy physics are usually more complicated than the simplest phenomenological inflationary models. Second, exploring  more  general models of inflation and identifying their specific observational features is a healthy procedure to prepare the interpretation of the future data.

At present, two types of extensions of the simplest scenarios have been mainly studied:
\bi \item models with non standard kinetic terms;
 \item models with multiple  scalar fields.
\ei
Of course, the two aspects can be combined and there exist  scenarios involving several scalar fields with non-standard kinetic terms, as we will see later. 
  
Among the scenarios involving several scalar fields, it is useful to distinguish three subclasses. The first, and oldest, category consists of models with {\it multiple inflatons}, i.e. models where several scalar fields play a dynamical role in the homogenous cosmological evolution during inflation.  
In the second category, one finds the {\it curvaton} scenarios. These models assume the existence, in addition to the inflaton,  of  a scalar field, which is light during inflation but does not participate to inflation {\it per se}. Its energy density, which decreases less quickly than radiation, becomes significant only after inflation. Its decay produces a second reheating, and its fluctuations are then imprinted in the curvature perturbation.

The final subclass regroups what we will name the {\it modulaton}  scenarios.
Like in the curvaton models, one assumes the presence of a light scalar field, the modulaton, which is subdominant during inflation but acquires some fluctuations. The fluctuations of the modulaton are transferred to the curvature perturbation because the cosmological evolution is governed by some parameter that depends on the modulaton. This parameter can be for instance the value of the inflaton at the end of inflation, or the coupling of the inflaton to other particles during the reheating. 
Of course, one can envisage even more complicated scenarios which combine several of these mechanisms.

\subsection{Generalized Lagrangians}
\index{generalized Lagrangians}
We now consider  multi-field models,  which can be described by an
action of the form
\beq
\label{P}
S =  \int d^4 x \sqrt{-g}\left[\frac{R}{16\pi G}   +   P(X^{IJ},\phi^K)\right] 
\eeq
where $P$ is an arbitrary function of $N$ scalar fields and of the  kinetic term
\beq
\label{X}
X^{IJ}=-\frac12 \nabla_\mu \phi^I  \nabla^\mu \phi^J.
\eeq

The relations obtained earlier for  the single field model can then be generalized, as we now show.
The energy-momentum tensor, derived from (\ref{P}),   is of the form
\beq
T^{\mu \nu} = P g^{\mu \nu} + P_{<IJ>}  \partial^\mu \phi^I \partial^\nu \phi^J\,,
\label{Tmunu}
\eeq
where $P_{<IJ>}$ denotes the partial derivative of $P$ with respect to $X^{IJ}$ (symmetrized with respect to the indices $I$ and $J$).
The equations of motion for the scalar fields, which can be seen as generalized Klein-Gordon equations, are obtained from the variation of the action with respect to $\phi^I$ and read
\beq
\nabla_{\mu} \left(  P_{<IJ>} \nabla^\mu \phi^J \right) + P_{,I} = 0\,.
\label{KG-general}
\eeq
where $P_{,I}$ denotes the partial derivative of $P$ with respect to $\phi^I$.

In a homogeneous  spacetime,  $X^{IJ}=\dot\phi^I\dot\phi^J/2$, and the energy-momentum tensor reduces to that of a perfect fluid
with pressure $P$ and energy density 
\beq
\rho=2  P_{<IJ>}  X^{IJ} - P\,.
\label{rho}
\eeq
 The evolution of the scale factor $a(t)$ is governed by the Friedmann equations, which can be written in the form
\beq
H^2 = \frac{1}{3 M_P^2} \left(2  P_{<IJ>}  X^{IJ} - P \right)\, , \qquad
\dot{H} = - X^{IJ} P_{<IJ>} /M_P^2\,.
\label{Friedmann2}
\eeq
The equations of motion for the scalar fields reduce to
\beq
\left( P_{<IJ>} + P_{<IL>,<JK>} \dot{\phi}^L \dot{\phi}^K  \right) \ddot{\phi}^J+ 
 \left( 3 H P_{<IJ>} + {P}_{<IJ>,K} \dot{\phi}^K\right)\dot{\phi}^J - P_{,I} = 0\, ,
\label{KG1}
\eeq
where $P_{<IL>,<JK>}$ denotes the (symmetrized) second derivative of $P$ with respect to $X^{IL}$ and $X^{JK}$.

The expansion of the action (\ref{P}) up to second order in the perturbations  is useful to obtain the classical equations of motion for the linear perturbations and to calculate the spectra of the primordial perturbations generated  during inflation, as we have seen earlier in the case of a single scalar field.  

 Working for convenience with the scalar field perturbations 
$Q^I$ defined in the spatially flat gauge, the
 second order action can be  written in the compact form \cite{lrst08b}
 \begin{eqnarray}
S_{(2)} &=& \frac{1}{2} \int {\rm d}t \, {\rm d}^3x \, a^3 \left[ 
\left(P_{<IJ>} + 2 P_{<MJ>,<IK>}X^{MK}\right) \dot{Q}^{I}\dot{Q}^{J}  
\right.
\cr
&& 
\qquad \left.
- P_{<IJ>} h^{ij} \partial_iQ^I \partial_jQ^J 
 - {\cal M}_{KL}Q^K Q^L 
 + 2 \,\Omega_{KI}Q^K \dot{Q}^I  \right] \, ,
 \label{S2}
\end{eqnarray}
where the mass matrix is
\begin{eqnarray}
{\cal M}_{KL} &=& - P_{,KL} + 3 X^{MN} P_{<NK>} P_{<ML>} 
+ \frac{1}{H} P_{<NL>}\dot{\phi}^N \left[ 2 P_{<IJ>,K} X^{IJ} - P_{,K} \right]
\nonumber
\\
& - & \frac{1}{H^2} X^{MN} P_{<NK>}P_{<ML>} \left[ X^{IJ} P_{<IJ>} + 2 
P_{<IJ>,<AB>} X^{IJ}X^{AB} \right] 
\nonumber
\\
& - & { \frac{1}{a^3}\frac{{\rm d}}{{\rm dt}} } \left( \frac{a^3}{H} P_{<AK>} P_{<LJ>} X^{AJ} \right)
\label{masssq}
\end{eqnarray}
and the mixing matrix is
\beq
\Omega_{KI} = \dot{\phi}^J P_{<IJ>,K} -
 \frac{2}{H} P_{<LK>} 
 %\left[  P_{<IJ>} X^{AJ} +  
 P_{<MJ>,<NI>} X^{LN}X^{MJ}\,.
%%+ \frac{\partial \hP_{MI}}{\partial X^{FL}}X^{AM}X^{FL} 
%\right]
\label{damping}
\eeq
This formalism is very general and, in the following, we will  consider two particular cases, which have often been studied in the literature.

%%%%%%%%%%%%%%%%%%%%%%%%
\subsection{Simple multi-inflaton scenarios}
%%%%%%%%%%%%%%%%%%%%%%%%

A more restrictive, although still very large, class of models consists of    multi-field scenarios governed by  a Lagrangian of the form 
\beq
P=G_{IJ}X^{IJ}-V=
-{1\over 2}\, 
G_{IJ}(\phi)
\, \partial^\mu\phi^I\partial_\mu\phi^J
-V(\phi)\, ,
\eeq
where the field metric $G_{IJ}$ can be non trivial
(also studied in  e.g.  \cite{Sasaki:1995aw,GrootNibbelink:2001qt,lr08}). 
It can then be shown that the  second-order action can be rewritten in the form  
\begin{eqnarray}
S_{(2)}&=& \half \int dt \,\dn{3}{x}\,    a^3 \left[ 
 G_{IJ}  \mathcal{D}_t Q^I \mathcal{D}_t Q^J -\frac{1}{a^2}G_{IJ} \partial_i Q^I \partial^i Q^J 
  -{\tilde M}_{IJ}Q^I Q^J  \right]\,,
\end{eqnarray}
with 
\begin{eqnarray}
{\tilde M}_{IJ}  &=& \mathcal{D}_I \mathcal{D}_J V - R_{IKLJ}\dot \p^K \dot
\p^L-\frac{1}{a^3}\mathcal{D}_t\left[\frac{a^3}{H}\dot \p_
I \dot \p_J\right]  \,,
\label{Interaction matrix}
\end{eqnarray}
and
where $ \mathcal{D}_I$ denotes the covariant derivative with respect to the field space metric $G_{IJ}$ (so that  $\mathcal{D}_I \mathcal{D}_J V=V_{,IJ}-\Gamma_{IJ}^K V_{,K}$ where the $\Gamma_{IJ}^K$ denote the Christoffel symbols of the metric $G_{IJ}$), while
$R_{IJKL}$ is the corresponding Riemann tensor and $ \mathcal{D}_t Q^I \equiv\dot Q^I+\Gamma^I_{JK}\dot\phi^I Q^K$.

It is now convenient, following the approach of \cite{Gordon:2000hv},   to introduce a particular direction in field space, which we will call the {\it instantaneous adiabatic} direction, defined by the unit vector tangent to the inflationary trajectory in field space,
\beq
\label{e_sigma}
e_\sigma^I=\frac{\dot{\phi}^I}{\sqrt{2X}}=\frac{\dot{\phi}^I}{\dot\s}\,,
\eeq
where we have introduced the  notation $X\equiv G_{IJ}X^{IJ}$ and $\dot\sigma\equiv \sqrt{2X}$.
% but it can be misleading as we have not defined any field $\sigma$.
%%%%%%%%%%%%%%%%%%%%%%%%%%%%%%%%%%%%%%%%%%%%%%%%%%%%
\begin{figure}
\begin{center}
\includegraphics[width=3.5in]{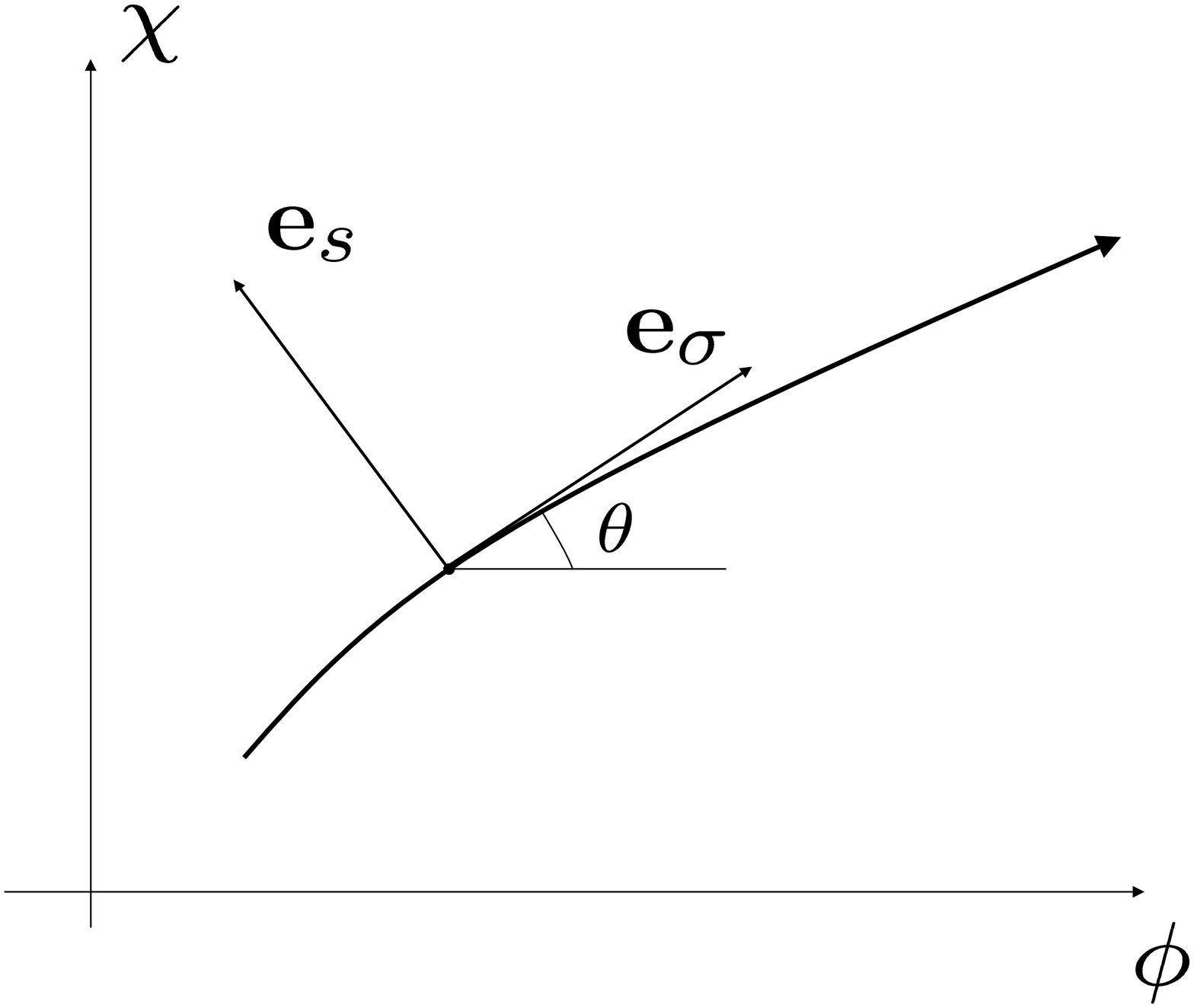}
\end{center}
\caption{Inflationary trajectory in a two-field model. The (instantaneous) adiabatic vector $e_\sigma$ is tangent to the trajectory while the (instantaneous) entropic vector $e_s$ is orthogonal to it. }
\end{figure}
%%%%%%%%%%%%%%%%%%%%%%%%%%%%%%%%%%%%%%%%%%%%%%%%%%%%
The  {\it instantaneous entropic} directions, which are  orthogonal to $e_\sigma^I$, span an hyperplane in field space. 

For simplicity, let us now concentrate on two-field scenarios, where   there is as single entropic degree of freedom. Defining
 the entropy vector $e^I_s$ as the  unit vector orthogonal to the adiabatic vector $e_\sigma^I$, i.e.
\beq
G_{IJ}e_s^I e_s^J=1, \qquad G_{IJ}e_s^I e_\sigma^J=0\, ,
\eeq
 the scalar field perturbations can  be uniquely  
decomposed into (instantaneous) adiabatic and entropic modes,
\beq
\label{decomposition}
Q^I=Q_\s e_\sigma^I+Q_s e^I_s\,.
\eeq

One can then derive the  equations of motion for the quantities $Q_\s$ and $Q_s$ from the second-order action. One finds~\cite{lr08}
\begin{eqnarray}
\label{Qsigma}
 \ddot{Q}_{\sigma}+ 3H
 \dot{Q}_{\sigma}+\left(\frac{k^2}{a^2}+\mu_{\s}^2\right)  Q_{\sigma} =
\Tdot{\left(\Xi Q_s\right)}
-\left(\frac{\dot H }{H}
+\frac{V_{,\sigma}}{\dot \s }\right)  \Xi\, Q_s\,, 
\end{eqnarray}
with
\beq
\Xi\equiv -\frac{2}{\dot \s}V_{,s} \,, \qquad
\mu_{\s}^2 \equiv  -\frac{\Tddot{(\dot\s/H)}}{\dot\s/H}-\left(3H+\frac{\Tdot{(\dot\s/H)}}{\dot\s/H}\right)
\frac{\Tdot{(\dot\s/H)}}{\dot\s/H}
\label{mu_sigma}\, ,
\eeq
where $V_{,\s}\equiv e_\s^I V_{,I}$ and  $V_{,s}\equiv e_s^I V_{,I}$.
The equation of motion for the entropy part is given by 
\beq
 \ddot{Q}_s+3H \dot{Q}_s+\left(\frac{ k^2}{a^2}+\mu_s^2\right)Q_s
  =-\Xi\left[\dot Q_{\sigma} -H\left(\frac{\dot \s^2}{2 H^2}+\frac{\ddot \s}{H \dot \s}\right)Q_{\sigma}\right] \, ,
\label{delta_s_all_scales}
\eeq
with
\beq
\mu_s^2\equiv V_{ss}+\half \dot \s^2 R-\frac{V_{,s}^2}{2X} \,,
\eeq
where
$R$ is the trace of the Ricci tensor on field space, i.e. the scalar curvature.

The adiabatic perturbation is naturally related to the comoving curvature perturbation (\ref{R}). Indeed, using the energy-momentum tensor 
(\ref{Tmunu}), with the property $\rho+P=2X$, which follows from (\ref{rho}), one finds that the comoving perturbation (\ref{R}) is given by 
 \beq
{\cal R}=\frac{H}{2X} \dot\phi_I Q^I=\frac{H}{\sqrt{2X}}Q_{\s}\,.
\eeq
Taking the time derivative of this expression and using the analog of (\ref{poisson}),
\beq
-2\frac{k^2}{a^2}\Psi=\delta\rho_c
= \sqrt{2X}\left[\dot Q_\s+\left(\frac{\dot H}{H}-\frac{\dot X}{2X}\right)Q_\s\right]+2V_{,s}Q_s
\label{drhom_Q}
\eeq
one finds
\beq
\label{Rdot}
\dot{ \mathcal{R}}=\frac{H}{\dot H}\frac{k^2}{a^2}\Psi-2\frac{H}{\dot\sigma^2}V_{,s}Q_s \, .
\eeq

By noting that the right hand side of Eq.~(\ref{delta_s_all_scales}) is proportional to $\dot\mathcal{R}$, one can rewrite the entropic equation of motion as
\beq
\ddot{Q}_s+3H\dot{Q}_s+\left(\frac{ k^2}{a^2}+\mu_s^2+\Xi^2\right) Q_s=-\frac{\dot\s}{\dot H}\Xi \frac{k^2}{a^2} \Psi\,.
\label{delta_s}
\eeq
When the spatial gradients can be neglected on large scales,  the above equation shows that the entropy perturbation $Q_s$ evolves independently of the adiabatic mode. In the same limit, the adiabatic mode is governed by a {\it first-order} equation
\beq
\dot{ \mathcal{R}} \approx \frac{H}{\dot \s}\, \Xi \, Q_s \quad {\rm or} \quad \dot{Q}_{\sigma}+\left(\frac{\dot{H}}{H}-\frac{\ddot{\sigma}}{\dot{\sigma}}\right) Q_{\sigma}-  \Xi \, Q_s \approx 0\,,
\label{R-large-scales}
\eeq
This implies that, in contrast with the entropy mode, the adiabatic mode is affected by the entropy on large scales, as soon as the {\it mixing parameter} $\Xi= -2 V_{,s}/\dot \s$ is non zero. When the field metric is flat, $G_{IJ}=\delta_{IJ}$, one can introduce the rotation angle $\theta$ between the initial basis and the adiabatic/entropy basis, which gives $\Xi=2\dot\theta$. In the case of a  field metric of the form 
\beq
G_{IJ}\, d\phi^I\, d\phi^J=d\phi^2+e^{2b(\phi)} d\chi^2\, ,
\label{metric_polar}
\eeq
investigated in  \cite{DiMarco:2002eb,Lalak:2007vi}, the coupling is now given by $\Xi=2\dot\theta+b'\dot\s \sin\theta$,  
where  the additional term simply comes from the non-trivial covariant derivative. 
Note that non-linear extensions of the adiabatic and entropic equations have been obtained in \cite{Langlois:2006vv,RenauxPetel:2008gi,Lehners:2009ja} (see also \cite{Rigopoulos:2005xx,Sasaki:1998ug} for other works on non-linear perturbations in multi-field inflation).

%%%%%%%%%%%%%%%%%%%%%%%%%%%%%%%%%%%%%%%%%%%%%%%%%%%%
\begin{figure}
\begin{center}
\includegraphics[width=4.5in]{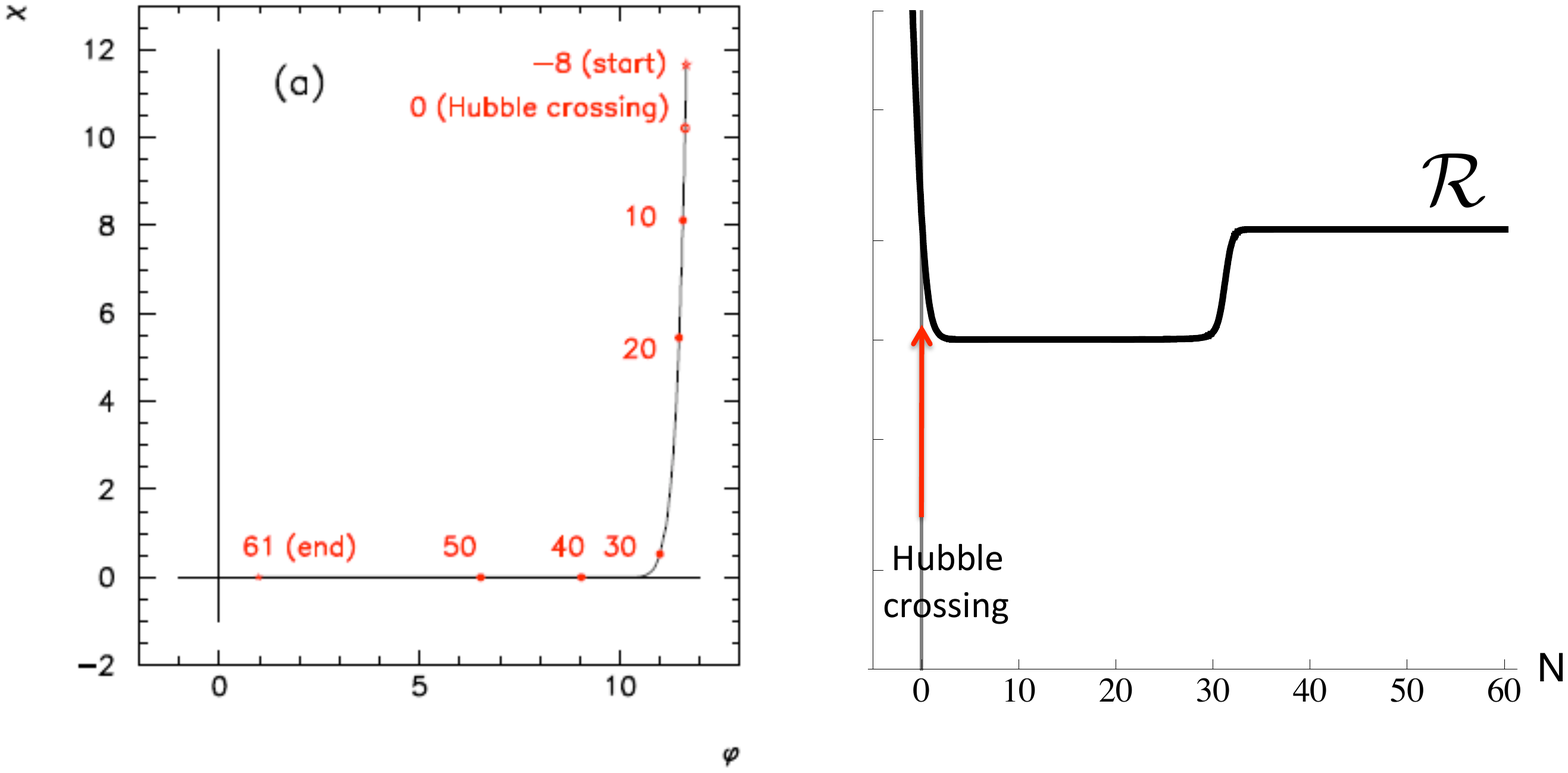}
\end{center}
\caption{In a double inflation model, with two different masses for the scalar fields, the inflationary trajectory is bent (left). This induces an evolution of the curvature perturbation, {\it after} Hubble crossing (right). Other examples can be found in \cite{Lalak:2007vi}.}
\label{double_inflation}
\end{figure}
%%%%%%%%%%%%%%%%%%%%%%%%%%%%%%%%%%%%%%%%%%%%%%%%%%%%

The above results show that a generic feature of multi-inflaton scenarios is that the curvature perturbation is not frozen after horizon crossing, like in single-field inflation, but can, instead, evolve on large scales as a consequence of the {\it transfer} of entropy perturbations into adiabatic perturbations, as illustrated on Fig.~\ref{double_inflation}. This property was pointed out originally in  \cite{sy} in the context of generalized gravity theories. 
As a consequence, it is  crucial, when working with a model involving several scalar fields during inflation, to identify all the light directions in field space and to evolve the curvature perturbation until any transfer from entropy into adiabatic modes has completely ceased (the transfer can even occur long after inflation, as is the case in the curvaton  scenario, which we will discuss later). 

As we have just seen, the instantaneous entropy perturbations can affect the evolution of the curvature perturbation during inflation, on large scales, but they could also survive the end of inflation and the reheating phase and therefore, cause the existence of  ``initial'' isocurvature perturbations, for instance between the  CDM and photon fluids, in the radiation era. Moreover, these isocurvature perturbations could be correlated with the ``initial'' adiabatic perturbations~\cite{Langlois:1999dw}, since part of the adiabatic perturbation can originate from an (instantaneous) entropy perturbation during inflation. We will discuss later the observational constraints on this possibility.

\subsection{K-inflation}
\index{K-inflation}
Let us now consider {\it single-field} inflation, but with a generalized Lagrangian
\beq
L=P(X,\phi), \qquad X\equiv-\partial_\mu\phi \partial^\mu\phi/2\, .
\eeq
This class of models, studied in   \cite{ArmendarizPicon:1999rj}, has been  called K-inflation because inflation can arise from the presence of non-standard kinetic terms, and not necessarily from a quasi-flat potential as in standard inflation. 

Linear perturbations have been investigated in \cite{Garriga:1999vw}. 
Here, they can simply be obtained from the single-field limit of (\ref{S2})
\begin{eqnarray}
S_{(2)} &=& \frac{1}{2} \int {\rm d}t \, {\rm d}^3x \, a^3 \left[ 
\left(P_{X} + 2 P_{XX}X\right) \dot{Q}^2  
- P_{X} h^{ij} \partial_iQ\partial_jQ 
\right.
\cr
&& 
\qquad \qquad \left.
 - {\cal M}Q^2 
 + 2 \,\Omega Q \dot{Q}  \right] \, ,
 \label{S2_k_inflation}
\end{eqnarray}
where $P_X\equiv \frac{\partial P}{\partial X}$ and $P_{XX}\equiv \frac{\partial^2 P}{\partial X^2}$.
The first line of the above action shows that the perturbations of the scalar field propagate with an effective sound speed given by
\index{sound speed}
\beq
\label{c_s}
c_s^2=\frac{P_X}{P_X+2XP_{XX}}\, ,
\eeq
which, in some models,  can be much smaller than the usual speed of light. 

Introducing the conformal time $\tau$ and the canonically normalized field 
\beq
v= \frac{a\sqrt{P_X}}{c_s}Q
\eeq
yields the action
\begin{eqnarray}
\label{S_v_k_inflation}
S_{(2)}&=&\frac{1}{2}\int {\rm d}\tau\,  {\rm d}^3x \left[
  v^{\prime\, 2}-c_s^2 (\partial v)^2 
+\frac{z''}{z} v^2\right]\, ,
\end{eqnarray}
with 
\beq
z=\frac{a\dot\phi\sqrt{P_X}}{c_s\, H}\, .
\eeq
In Fourier space, this leads to the equation of motion
\beq
v''+\left(k^2c_s^2-\frac{z''}{z}\right)v=0,
\eeq
\index{sound horizon}
where one notes the presence of $c_s^2$ multiplying $k^2$. As a consequence, the fluctuations are amplified at  {\it sound horizon} crossing, 
i.e. when $k c_s\sim aH$, and not at Hubble radius crossing as in the standard case (the two of course coincide for $c_s\simeq c$). 

Assuming  a slow variation of the Hubble parameter $H$ and of the sound speed $c_s$, one can use the approximation $z''/z\simeq 2/\tau^2$ and the solution 
corresponding to the vacuum on small scales is given by
\beq
v=\frac{1}{\sqrt{2kc_s}}e^{-ikc_s\tau}\left(1-\frac{i}{kc_s\tau}\right).
\eeq  
This expression differs from (\ref{u_k}) only by the presence of $c_s$.  

One can then proceed exactly as in the standard case to obtain the power spectrum of the scalar field fluctuations
\beq
{\cal P}_{Q}\simeq\frac{H^2}{4\pi^2c_s P_X }
\eeq
and 
the power spectrum of the curvature perturbation 
 \beq
{\cal P}_{\cal R_*}=\frac{k^3}{2\pi^2}\frac{|v_{\s\, k}|^2}{z^2}\simeq\frac{H^4}{4\pi^2  \dot\s^2 }=\frac{H^2}{8\pi^2 \epsilon c_s }\,,
\label{power-spectrum-R}
\eeq
where $\epsilon=-\dot H / H^2\,$.

\subsection{A specific example: multi-field DBI inflation}
\index{DBI inflation}
The two previous subsections have illustrated {\it separately} the consequences of multiple inflatons, on the one hand, and of non-standard kinetic terms, on the other hand. Here, these two aspects will be naturally combined in  a category of models motivated by string theory, where inflation is due to  the motion  of a  $D3$-brane in an internal  six-dimensional compact space. 

 The dynamics of the brane, with tension $T_3$, is governed by the Dirac-Born-Infeld Lagrangian (we ignore here the dilaton and the various form fields, but they can be included, as in \cite{Langlois:2009ej})
\beq
\label{L}
L_{\rm DBI} =- T_3\sqrt{-\det{\gamma_{\mu \nu}}}
\eeq
which depends on the determinant of the induced metric on the 3-brane,
\beq
\gamma_{\mu \nu} =H_{AB}\,  \partial_\mu Y_{\rm (b)}^A \partial_\nu Y_{\rm (b)}^B 
\label{DBIzero}\, ,
\eeq
where $H_{AB}$ is metric of the compactified  10-dimensional spacetime, assumed to be of the form
\beq
H_{AB} \, dY^A \, dY^B=h^{-1/2}(y^K)\,g_{\mu \nu}dx^\mu dx^\nu + h^{1/2}(y^K)\, G_{IJ}(y^K)\, dy^I dy^J \, , 
\eeq
 and  $Y_{\rm (b)}^A(x^\mu)=(x^\mu, \psi^I(x^{\mu}))$, with $\mu=1\dots 3$ and $I=1\dots 6$, defines the  brane embedding.

After using the  rescalings  $\phi^I\equiv\sqrt{T_3} Y^I$ and $f=h/T_3$, one ends up with a Lagrangian of the form
\beq
P= -\frac{1}{f(\bfphi^I)}\left(\sqrt{{\cal D}}-1\right) -V(\bfphi^I)
\label{DD}
\eeq
with
\begin{eqnarray}
\label{def_explicit}
{\cal D} &\equiv&  \det(\delta^{\mu}_{\nu}+f \, G_{IJ}\partial^{\mu} \phi^I \partial_{\nu} \phi^J )\cr
&=&1-2f G_{IJ}X^{IJ}+4f^2 X^{[I}_IX_J^{J]} -8f^3 X^{[I}_IX_J^{J} X_K^{K]}+16f^4 X^{[I}_IX_J^{J} X_K^{K}X_L^{L]}
\end{eqnarray}
where the field indices are lowered by the field metric $G_{IJ}$, i.e. the metric of the internal compact space,  and the brackets denote antisymmetrization of the indices. 
A  potential term, which arises from the brane's interactions with bulk fields or other branes, is also included. 

Assuming that the brane is moving in a  conical geometry, many works  have concentrated on the purely radial dynamics of the brane, while ignoring the angular directions. 
The  effective action then reduces to
\beq
\label{action_DBI}
S=\int d^4x \sqrt{-g}\left[- \frac{1}{f}\left(\sqrt{1+f  \, \partial_\mu\phi\partial^\mu \phi} -1\right)
-V(\phi)\right]\, .
\eeq
If $f\dot\phi^2\ll 1$, one can expand the square root in the Lagrangian and one recovers the usual kinetic term familiar to slow-roll inflation. 

This Lagrangian also leads to  another type of inflation, called DBI inflation \cite{st,ast}, in the ``relativistic'' limit
\beq
1-f  \, \dot\phi^2\ll 1 \quad \Leftrightarrow \quad |\dot\phi|\simeq 1/\sqrt{f}\, .
\eeq
Indeed, using (\ref{Friedmann2}), one can check that it is possible to obtain $\epsilon\equiv -\dot H/H^2\ll 1$ in this limit, provided $V\gg 1/f c_s$. 
An interesting property of DBI inflation is that the potential can be rather steep, in contrast with  standard slow-roll inflation.

The Lagrangian in (\ref{action_DBI}) is of the form $P(X,\phi)$, discussed in the previous subsection, with
\beq
P(X,\phi)=-\frac{1}{f(\phi)}\left(\sqrt{1-2fX}-1\right)-V(\phi),
\eeq
and therefore, using (\ref{c_s}), 
\beq
c_s=\sqrt{1-2fX}=\frac{1}{P_X}.
\eeq

If  the angular directions are relevant, the above single field simplification is not valid and one must work in  a multi-field framework with the  Lagrangian (\ref{DD}). The perturbations generated by such a  scenario have been studied in detail in \cite{lrst08a} and 
we now summarize the main results. 

After decomposing the perturbations into adiabatic and entropy modes, one finds that the single field results apply to the adiabatic mode, so that its spectrum at {\it sound horizon crossing} is given by 
\beq
\label{power_sigma}
{\cal P}_{Q_\s*}\simeq\frac{H^2}{4\pi^2 }
\eeq
(the subscript $*$ here indicates that the corresponding quantity is evaluated at sound horizon crossing $k c_s=aH$).

As for the (canonically normalized) entropy mode, $v_s\equiv (a/\sqrt{c_s})Q_s$,  its evolution is governed by the equation 
\begin{eqnarray}
v_{s}''+\xi  v_{\s}'+\left(k^2 c_s^2- \frac{\alpha''}{\alpha}\right) v_{s} =0\,, \qquad \alpha\equiv \frac{{a}}{\sqrt{c_s}}
\label{eq_v_s}
\end{eqnarray}
 where we have neglected a possible coupling with the adiabatic mode and assumed that the effective mass of the entropy mode is small with respect to $H$. $v_s$ has thus the same spectrum as $v_\s$, but since the normalization coefficients in front of the adiabatic and entropy modes differ, one finds that the spectrum for the fluctuations along the entropy direction in field space, is given by
\beq
 {\cal P}_{Q_s*}\simeq\frac{H^2}{4\pi^2 c_s^2},
\eeq
which shows that, for small $c_s$, the entropic modes are {\it amplified} with respect to the adiabatic modes:
\beq
Q_{s*}\simeq \frac{Q_{\sigma*}}{c_s}.
\eeq

Since we are in a multi-field scenario,  the curvature perturbation can be modified, after sound horizon crossing,  if there 
is a transfer  from the entropic to the adiabatic modes, as we saw earlier. The final curvature perturbation can be formally written as
 \beq
 \R=\R_*+T_{ {\cal R}  {\cal S} } \cal S_*\, ,
 \eeq
  where,  for convenience, we have introduced  the {\it rescaled} entropy perturbation
  \beq
{\cal S}=c_s\frac{H}{\dot \s}Q_{s}\, ,
\label{S}
\eeq
defined such that its power spectrum
at sound horizon crossing is the same as that of the curvature perturbation, i.e.
 ${\cal P}_{\cal S_*}={\cal P}_{\cal R_*}$.
The final curvature power-spectrum is thus given by
\beq
{\cal P}_{\cal R}=(1+T_{{\cal R} {\cal S}}^2) {\cal P}_{\cal R_{*}}=\frac{{\cal P}_{\cal R_{*}}}{{\rm cos^2} \Theta}\, ,
\label{observed-spectrum}
\eeq
where we have introduced  the ``transfer angle'' $\Theta$ defined by
\beq
%{\rm cos \Delta}
{\sin} \Theta =\frac{T_{ {\cal R}  {\cal S} }}{\sqrt{1+T^2_{ {\cal R}  {\cal S} }}}
\label{correlation-result}
\eeq
 (so that $\Theta=0$ if there is no transfer and $|\Theta|=\pi/2$ if the final curvature perturbation is mostly of entropic origin). 

The power spectrum for the tensor modes is still governed by the transition at {\it Hubble radius} and its amplitude, given by (\ref{power_T}),
is unchanged. The tensor to scalar ratio is thus 
\beq
r \equiv \frac{{\cal P}_{T}}{{\cal P}_{\cal R}}=16\, \epsilon\, c_s {\cos^2} \Theta\, .
\eeq
Interestingly this expression combines the result of $k$-inflation~\cite{Garriga:1999vw}, where the ratio is suppressed by a small sound speed $c_s$, and that of multi-field inflation with standard kinetic terms~\cite{Bartolo:2001rt}, where the ratio is suppressed by a large transfer from entropy to adiabatic modes.

\subsection{The curvaton scenario}
\index{curvaton}
\def\curv{\sigma}
\def\zetar{\zeta_{\rm r}}
% in case ever want to distinguish \zeta_rad and \zeta_r, could redefine here
\def\zetarad{\zeta_{\rm r}}
\def\i{{\rm inf}}

The transfer from entropy  into adiabatic perturbations can occur during inflation, as we have seen in scenarios with multiple inflatons, but it can also take place long after the end of inflation. A much studied example of this possibility is the curvaton scenario~\cite{Enqvist:2001zp,Lyth:2001nq,Moroi:2001ct} (see also \cite{Linde:1996gt}). 

 The curvaton is a weakly coupled scalar field,
$\curv$,
which is light relative to the Hubble
rate during inflation, and hence acquires perturbations during inflation, with an almost
scale-invariant power spectrum 
\beq
\label{spectrum_curvaton}
{\cal P}_{\delta\curv}=\left(\frac{H}{2\pi}\right)^2\, ,
\eeq
where the curvaton perturbation is defined here in the flat gauge, i.e. $\delta\curv=Q_\curv$.

%%%%%%%%%%%%%%%%%%%%%%%%%%%%%%%%%%%%%%%%%%%%%%%%%%%%
\begin{figure}
\begin{center}
\includegraphics[width=3.5in]{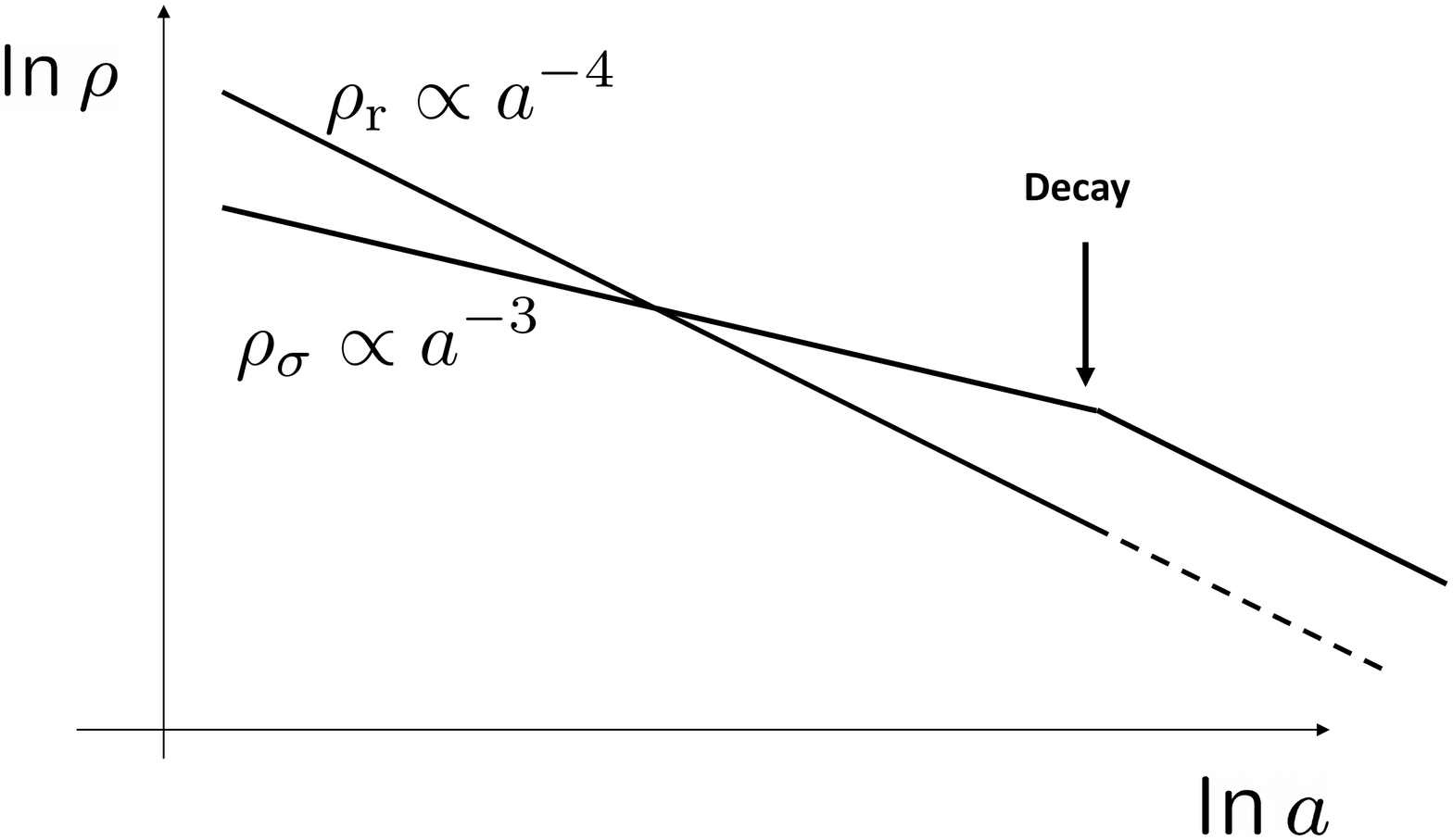}
\end{center}
\caption{Evolution of the energy density of the radiation, $\rho_r$, produced by the inflaton and of the energy density of the curvaton, $\rho_\sigma$, before and after the curvaton decay.}
\label{curvaton}
\end{figure}
%%%%%%%%%%%%%%%%%%%%%%%%%%%%%%%%%%%%%%%%%%%%%%%%%%%%

 After inflation, 
the Hubble rate drops and, eventually, the curvaton becomes
non-relativistic so that its energy density grows relatively to that of 
radiation, until it represents a significant fraction of the
total energy density, $\Omega_\curv\equiv\bar\rho_\curv/\bar\rho$,
before it finally decays (see Fig.~\ref{curvaton}). Hence the initial curvaton field perturbations
on large scales can give rise to a primordial density perturbation
after the decay of the curvaton.

Before it decays, the non-relativistic curvaton (with mass $m\gg H$)
 behaves effectively as a pressureless, non-interacting fluid with
energy density
 \begin{equation}
\rho_\curv = m^2 \curv^2 \;,
 \end{equation}
where $\curv$ is the rms amplitude of the curvaton field.
The corresponding perturbations are characterized, using (\ref{zeta}) and (\ref{spectrum_curvaton}), by
\beq
\zeta_\curv=\left(\frac{\delta\rho_\curv}{3\rho_\curv}\right)_{\rm flat}=\frac23\frac{\delta\curv}{\curv}
\quad \Rightarrow {\cal P}_{\zeta_\curv}\simeq \frac{H^2}{9\pi^2 \curv^2}\, .
\eeq
When the curvaton decays into radiation, its perturbations are converted into perturbations of the resulting radiation
fluid. The subsequent perturbation is described by
 \begin{equation}
 \label{zeta_post_decay}
 \zeta = r_\curv \zeta_\curv +(1-r_\curv)\zeta_\i\,, \quad r_\curv \equiv \frac{3\Omega_{\curv,{\rm decay}}}{4-\Omega_{\curv,{\rm
decay}}}\, .
 \end{equation}
This implies that
the power spectrum for the primordial adiabatic perturbation
$\zetarad$ can be expressed as
 \beq
 \label{finalzetar}
  {\cal P}_{\zeta}={\cal P}_{\zeta_\i}+r_\curv^2{\cal P}_{\zeta_\curv} \;.
 \eeq
where ${\cal P}_{\zeta_\i}$ is the spectrum of perturbations generated directly by the inflaton fluctuations.
 
 In the case of
single field inflation, $ {\cal P}_{\zeta_\i}$ is given in (\ref{power_S}) and one can rewrite the total power spectrum as 
we have
 \beq
{\cal P}_{\zeta}=(1+\lambda){\cal P}_{\zeta_\i} \, ,\qquad 
\lambda \equiv \frac{8}{9}r_\curv^2\epsilon_*
\left(\frac{\curv_*}{\mP}\right)^{-2}
\label{lambda} 
\eeq
The limit  $\lambda\gg 1$ corresponds to  the original  curvaton scenario where
the inflaton perturbations are negligible: since $r_\curv$ and $\epsilon_*$
are bounded by $1$, this requires $\curv_*\ll \mP$. 

A value of
$\lambda$ of order $1$ or smaller is possible if $r_\curv$ or $\epsilon_*$
are sufficiently small and/or $\curv_*$ is of the order of $\mP$. In the latter case the curvaton starts to oscillate at about the same time
as it decays and cannot be described as a dust field.   A more refined treatment~\cite{Langlois:2004nn}
 shows that the curvature perturbation due to the inflaton and curvaton perturbations is given
\beq
{\cal R}=-{V\over \mP^2 V'}\delta\phi
-{3\over 2}f(\sigma_*){\delta\sigma_*\over \mP},
\eeq
where the function $f(\sigma_*)$ interpolates between the limiting situations of a pure curvaton and of 
a secondary inflaton, 
\beq
f(\curv_*)\simeq
\left\{
\begin{array}{cc}
 \frac49\frac{\mP}{\curv_*}, &\quad  \curv_*\ll \mP
 \\
 \\
\frac {\curv_*}{ 3 \mP}, & \curv_*\gg \mP.
 \end{array}
 \right.
\eeq

Interestingly, the curvaton scenario can also produce entropic, or isocurvature,  perturbations~\cite{Lyth:2002my}. It can produce a CDM isocurvature perturbation  if the CDM is created before the curvaton decay and thus inherits the perturbations of the inflaton so that $S_{\rm cdm}=3(\zeta_{\rm inf}- \zeta_r)$; or, on the contrary, if the CDM is created by the curvaton decay, in which case $S_{\rm cdm}=3(\zeta_\sigma-\zeta_r)$. Similarly, baryon isocurvature perturbations can be generated if the baryon asymmetry exists before the curvaton decay.

\subsection{Modulaton}
\index{modulaton}
\def\mod{\sigma}
In the curvaton scenario, the curvaton dominates the energy density of the Universe at some epoch in order to give the main contribution to the primordial perturbations.
Alternatively,  one can also envisage scenarios where the primordial perturbations are due to the perturbations of a scalar field, which has never dominated the matter content of the universe but has played a crucial r\^ole  during some  cosmological transition. We will name this field a {\it modulaton}. 

\index{modulated reheating}
The best example is the {\it modulated reheating} scenario~\cite{Dvali:2003em,Kofman:2003nx}  where the decay rate of the inflaton, $\Gamma$,  depends on a modulaton $\mod$, which has acquired classical fluctuations during inflation. The decay rate is thus slightly different from one super-Hubble patch to another, which generates a curvature perturbation. 

A simple way to quantify this effect is to compute the number of e-folds between some initial time $t_i$ during inflation, when the scale of interest crossed out the Hubble radius, and some final time $t_f$. The curvature perturbation is then directly related  to the fluctuations of the number of e-folds, as we discussed at the beginning of Section 5.

For simplicity, we will assume that,  just after the end of inflation at time $t_e$, the inflaton behaves like pressureless matter (as is the case for a quadratic potential) until it decays instantaneously at the time $t_d$
 characterized by $H_d=\Gamma$. 
 At the decay, the energy density is thus $\rho_d=\rho_e\exp[-3(N_d-N_e)]$ and is transferred into radiation, so that, at time $t_f$, one gets
 \beq
 \label{rho_f_modulaton}
 \rho_f=\rho_de^{-4(N_f-N_d)}=\rho_e e^{-3(N_f-N_e)-(N_f-N_d)}.
 \eeq
 Using the relation  $\Gamma=H_d=H_f\exp[2(N_f-N_d)]$ to eliminate $(N_f-N_d)$ in (\ref{rho_f_modulaton}), we finally obtain 
 \beq 
 N_f=N_e-\frac13\ln\frac{\rho_f}{\rho_e}-\frac16\ln\frac{\Gamma}{H_f}\, .
 \eeq
 If one ignores the inflaton fluctuations, the final curvature perturbation is therefore 
 \beq
 \zeta=N_{,\mod}\delta\mod_*=-\frac16\frac{\Gamma_{,\mod}}{\Gamma}\delta\mod_*,
\eeq
which yields the curvature power spectrum
\beq
\P_\zeta=\frac{1}{36}\left(\frac{\Gamma_{,\mod}}{\Gamma}\right)^2\left(\frac{H_*}{2\pi}\right)^2\, .
\eeq
The dependence on the modulaton can alternatively show up in the mass of the particles created by the decay of the inflaton~\cite{Dvali:2003ar,Vernizzi:2003vs}.

The modulaton can also affect the cosmological evolution {\it during} inflation, as in the {\it modulated trapping} scenario~\cite{Langlois:2009jp}, which is based on the resonant 
production of particles during inflation~\cite{Chung:1999ve} (see also \cite{Barnaby:2009mc,Green:2009ds} for other recent scenarios based on particle production). If the inflaton is coupled to some particles, whose effective mass becomes zero for a critical value of the inflaton, then there will be a burst of production of these particles when the inflaton crosses the critical value. These particles will be quickly diluted but they will slow down the inflaton. This effect, which increases the number of e-folds until the end of inflation, can depend on a modulaton, for example via the coupling between the inflaton and the particles, and a significant curvature perturbation might be generated (see \cite{Langlois:2009jp} for details).

\subsection{``Initial '' adiabatic and entropic perturbations}
In contrast with single field inflation, multi-field inflation can  generate isocurvature ''initial'' perturbations in the radiation era. Note that this is only a possibility but not a necessity: purely adiabatic initial conditions are perfectly compatible with multi-field scenarios. 
 
The CMB is a powerful way to study isocurvature perturbations because  
 (primordial) adiabatic and isocurvature perturbations 
produce very distinctive 
features in the CMB anisotropies.   
On large angular scales, 
one can show for instance that \cite{Langlois:1999dw}
\beq
\frac{\delta T}{T}\simeq \frac{1}{5}
\left({\cal R}
-2 S\right).
\eeq
On smaller angular scales, an adiabatic initial perturbation  generates a cosine 
oscillatory mode in the photon-baryon fluid, leading to 
an acoustic peak at $\ell\simeq 220$ (for a flat universe), whereas a
  pure isocurvature initial perturbation generates a sine oscillatory 
mode resulting in a first peak at $\ell \simeq 330$. 
The unambiguous observation of the first peak at $\ell\simeq 220$ has 
eliminated the possibility of a dominant isocurvature perturbation. 
The recent observation by WMAP of the CMB polarization has also confirmed 
that the initial perturbation is mainly an adiabatic mode. But this does not 
exclude the presence of a subdominant isocurvature contribution, which 
could be detected in  future high-precision experiments such as Planck.

The combined impact of adiabatic and entropic perturbations crucially depends on their correlation \cite{Langlois:1999dw,Langlois:2000ar}
\beq
\beta=
\frac{ {\cal P}_{_{{S},{\cal R}}}}
{\sqrt{{\cal P}_{_{{S}}}{\cal
  P}_{_{\cal R}}}}.
\eeq

Parametrizing the relative amplitude between the two types of perturbations  by a coefficient $\alpha$, 
\beq
\frac{{\cal P}_{{S}}}{{\cal
  P}_{\cal R}}\equiv \frac{\alpha}{1-\alpha}\, ,
\eeq
the WMAP5 data \cite{wmap5} yield the following constraints  on the entropy contribution 
\beq
\beta=0:\ \alpha_0<0.067 \ (95\%\, {\rm C.L.})  \qquad \beta=-1: \ \alpha_{-1}<0.0037 \ (95\%\, {\rm C.L.})
\eeq
  in the 
uncorrelated case ($\beta=0$) and 
in the totally anti-correlated case 
($\beta=-1$), respectively.

%%%%%%%%%%%%%%%%%%%%%%
%%%%%%%%%%%%%%%%%%%%%%
\section{Primordial non-Gaussianities}
%%%%%%%%%%%%%%%%%%%%%%
%%%%%%%%%%%%%%%%%%%%%%%
\index{non-Gaussianities}
One of the most promising probes of the early Universe, which has been investigately very actively in the last few years, is the non-Gaussianity of the primordial perturbations (see \cite{Bartolo:2004if} for a review, but the field has grown considerably in the last few years). Whereas the simplest models of inflation, based on a slow-rolling single field with  standard kinetic term, generate undetectable levels of non-Gaussianity~\cite{Maldacena:2002vr,Acquaviva:2002ud}, a significant amount of non-Gaussianity  can be produced in scenarios with i) non-standard kinetic terms; ii) multiple fields; iii) a non standard vacuum; iv) a non slow-roll evolution. We will discuss in this section the first two possibilities.

\subsection{Higher order correlation functions}
The most used estimate of non-Gaussianity is the bispectrum defined, in Fourier space, by 
\index{bispectrum}
\beq
\langle
 \zeta_{\bk_1} \zeta_{\bk_2} \zeta_{\bk_3} 
\rangle \equiv (2 \pi)^3
\delta^{(3)}(\sum_i \bk_i) 
B_\zeta (k_1,k_2,k_3)\, ,
\eeq
where the Fourier modes are defined by 
\beq
\label{fourier2}
\zeta_\bk=\int d^3\bx \ e^{-i \bk\cdot\bx} \, \zeta(\bx)\, .
\eeq
Equivalently, one often uses the so-called $f_{\rm NL}$ parameter, which can be defined in general by
\beq
\label{f_NL}
B_\zeta (k_1,k_2,k_3)\equiv
\frac{6}{5}
f_{\rm NL} (k_1,k_2,k_3)\left[P_\zeta(k_1)P_\zeta(k_2)+P_\zeta(k_2)P_\zeta(k_3)+P_\zeta(k_3)P_\zeta(k_1)\right]\, ,
\eeq
where  $P_\zeta$ is the power spectrum\footnote{In this section on non-Gaussianities, we have followed the recent literature and adopted the definition (\ref{fourier2}) for the Fourier modes, which differs slightly from our convention (\ref{Fourier_quantum}) of the previous chapters. This changes the expression of the power spectrum, but the quantiy $\P(k)$ is the same in the two conventions.}
defined by
\beq
\langle
\zeta_{\bk_1} \zeta_{\bk_2} \rangle = (2
\pi)^3 \delta^{(3)} (\bk_1 +
\bk_2) \,  P (k_1) \, .
\eeq

The $f_{\rm NL}$ parameter was initially introduced in \cite{Komatsu:2001rj} for a very specific type of non-Gaussianity characterized by
\beq
\zeta(\x)=\zeta_{G}(\x)+\frac35f_{\rm NL}\zeta_{G}^2(\x)\, ,
\eeq
in the physical space, 
where $\zeta_{G}$ is Gaussian and the factor $3/5$ appears because $f_{\rm NL}$ was originally defined with respect to the gravitational potential $\Phi=(3/5)\zeta$, instead of $\zeta$. In this particular case, $f_{\rm NL} $, as defined in (\ref{f_NL}),  is independent of the vectors $\bk_i$. 
In general, $f_{\rm NL} $ is a function of  the norm of the three  vectors $\bk_i$ (which define a triangle in Fourier space since they are constrained by $\bk_1+\bk_2+\bk_3=0$ as a consequence of homogeneity),  and the ``shape'' of the three-point function is an important characterization of how non-Gaussianity was generated \cite{Babich:2004gb}.

In the context of multi-field inflation, the
so-called $\delta N$-formalism~\cite{Starobinsky:1986fxa, Sasaki:1995aw} is particularly
useful to evaluate the primordial non-Gaussianity generated on large scales
\cite{Lyth:2005fi}. The idea is to describe, on scales larger than the Hubble radius, 
the non-linear evolution of perturbations generated during  inflation in
terms of the perturbed expansion from an initial flat  hypersurface
(usually
taken at Hubble crossing
during inflation) up to a final uniform-density
hypersurface (usually during the radiation-dominated era).
Using the Taylor expansion of the number of e-folds given as a function of the initial values of the scalar fields, 
\beq
\label{taylor}
\zeta \simeq  \sum_I N_{,I} \delta \varphi_*^I + \frac{1}{2}
\sum_{IJ} N_{,IJ} \delta \varphi_*^I \delta \varphi_*^J
\eeq
one finds~\cite{Lyth:2005fi,Seery:2005gb}, in Fourier space, 
\begin{eqnarray}
\langle \zeta_{\bk_1} \zeta_{\bk_2} \zeta_{\bk_3} \rangle &=&
\sum_{IJK} N_{,I} N_{,J} N_{,K} \langle \delta \varphi^I_{\bk_1}
\delta \varphi^J_{\bk_2} \delta \varphi^K_{\bk_3}\rangle + \nonumber
\\ &&  \frac{1}{2} \sum_{IJKL} N_{,I} N_{,J} N_{,KL} \langle \delta
\varphi^I_{\bk_1} \delta \varphi^J_{\bk_2} (\delta \varphi^K \star
\delta \varphi^L)_{\bk_3}\rangle
+{\rm perms}. \nonumber \\ &&
\label{tpf}
\end{eqnarray}

The first term on the right hand side  corresponds  to  non-Gaussianities arising from nonvanishing   three-point function(s) of the scalar field(s). This is  the case for models with non-standard kinetic terms~\cite{Creminelli:2003iq,Seery:2005wm,Chen:2006nt}, leading to  a specific shape of non-Gaussianities, usually called {\it equilateral}, where the dominant contribution comes from configurations with  three wavevectors of similar length $k_1\sim k_2\sim k_3$. 

The terms appearing in second line  of (\ref{tpf}) can also lead to sizable non-Gaussianities. Indeed, 
 substituting
\beq
\langle
\delta\varphi^I_{\bk_1} \delta\varphi^J_{\bk_2} \rangle = (2
\pi)^3 \delta_{IJ} \delta^{(3)} (\bk_1 +
\bk_2)  \frac{2 \pi^2}{k_1^3} \PP_* (k_1) , \qquad \PP_*(k)  \equiv \frac{H_*^2}{4
\pi^2},
\eeq
in (\ref{tpf}), 
one gets
\beq
\label{f_local}
\frac{6}{5}f_{\rm NL} = 
  \frac{
N_{,I} N_{,J} N^{,IJ}}{( N_{,K}N^{,K})^2}\, ,
\eeq
where we use Einstein's summation convention for the field indices, which are raised with $\delta^{IJ}$.  
This corresponds to another type of non-Gaussianity, usually called {\it local} or {\it squeezed}, for which the dominant contribution comes from configurations where the three wavevectors form a squeezed triangle.

The present observational constraints \cite{wmap5} are 
\beq
-9< f_{NL}^{\rm (local)}<111 \quad (95\% \, {\rm CL}), \qquad -151< f_{NL}^{\rm (equil)}<253 \quad (95\%\,  {\rm CL}),
\eeq
 for  the local non-linear coupling parameter
and the equilateral non-linear coupling parameter, respectively.

Extending  the Taylor expansion (\ref{taylor}) up to third order, one can compute in a similar way the trispectrum~\cite{Seery:2006vu}, i.e. the Fourier transform of the connected four-point function defined by
\index{trispectrum}
\beq
\langle \zeta_{\bk_1} \zeta_{\bk_2} \zeta_{\bk_3} \zeta_{\bk_4} \rangle_{c} \equiv (2 \pi)^3
\delta^{(3)}(\sum_i \bk_i) 
T_\zeta (\bk_1, \bk_2, \bk_3, \bk_4)\, .
\eeq
Assuming the scalar field perturbations to be quasi-Gaussian, the trispectrum can be written in the form~\cite{Byrnes:2006vq}
\begin{eqnarray}
\label{trispectrum}
T_\zeta (\bk_1, \bk_2, \bk_3, \bk_4)&=&\tau_{\rm NL}\left[P(k_{13})P (k_3)P(k_4)+ 11 \ {\rm perms}\right]
\\
&& +\frac{54}{25} g_{\rm NL}\left[P(k_2)P(k_3)P(k_4)+3\ {\rm perms}\right],
\end{eqnarray}
with 
 \beq
  \tau_{\rm NL}= \frac{N_{IJ}N^{IK}N^JN_K}{(N_LN^L)^3}, \qquad
  g_{\rm NL}=\frac{25}{54}\frac{N_{IJK}N^IN^J N^K}{(N_LN^L)^3}
  \eeq
and where $k_{13}\equiv\left|\bf {k}_1+\bf {k}_3\right|$.

\subsection{A few examples}
It is not always easy to obtain significant non-Gaussianities even in inflationary models with several inflatons (see e.g. \cite{Bernardeau:2002jy,Bernardeau:2002jf,Seery:2005gb,Vernizzi:2006ve,Battefeld:2006sz,Gao:2008dt,Byrnes:2009qy}). 
Local non-Gaussianity can also be generated  at the end of inflation~\cite{Enqvist:2004ey,Lyth:2005qk,Sasaki:2008uc}.
 Below, we  discuss in more details the non-Gaussianities generated, first, in the curvaton scenario and, then, in multi-field DBI inflation.

\subsubsection{Curvaton}
In scenarios with a curvaton (or a modulaton), the total number of e-folds can be written as the sum of  two contributions:  one from the inflaton field $\phi$ and the other from the curvaton/modulaton $\sigma$. In the case of standard slow-roll inflation, the second derivatives with respect to $\phi$ are negligible and  (\ref{f_local}) reduces to 
\beq
\frac{6}{5}f_{\rm NL} =  \frac{N_\sigma^2\, N_{\sigma\sigma}}{(N_\phi^2+N_\sigma^2)^2}=  \frac{N_{\sigma\sigma}}{N_\sigma^2(1+\lambda^{-1})^2},
\eeq
where we have introduced the parameter $\lambda\equiv N_\sigma^2/N_\phi^2$, which represents the ratio of the contribution of $\sigma$ with that of the inflaton  in the power spectrum (see (\ref{lambda}) for the curvaton). 

For the curvaton,  Eq.~(\ref{zeta_post_decay}) tell us that $N_\sigma=2r_\curv/3\curv$ and  the extension of this equation to second order yields
\beq
N_{\sigma\sigma}=\frac{4r_\curv}{9\curv^2}\left(\frac32-2r_\curv -r_\curv^2\right),
\eeq
which leads to a local non-Gaussianity characterized by 
 \beq
\frac65 f_{\rm NL} =
\frac{1}{r_\curv}\frac{\left(\frac32-2r_\curv-r_\curv^2\right)}{(1+\lambda^{-1})^2}
\label{f_ad}\, .
 \eeq
Non-Gaussianities are thus
significant when the curvaton decays well before it dominates,
$r_\curv\ll 1$.
When $\lambda\gg 1$ and the perturbations from inflation are
negligible, one recovers the standard curvaton result \cite{Lyth:2002my}.

Note however that $f_{\rm NL}$ does not grow indefinitely as  $r_\curv$ becomes small because both $r_\curv$ and $\lambda$ depend on the curvaton
expectation value $\curv_*$. Indeed, 
substituting  $r_\curv\sim
(\curv_*/\mP)^2/\sqrt{\Gamma_\curv/m_\curv}$ (valid in the limit
$r\ll 1$), where $\Gamma_\curv$ is the decay
rate of the curvaton, into the definition (\ref{lambda}), one sees
that $\lambda$ is proportional to $\curv_*^2$, like $r_\sigma$. One thus finds~\cite{Ichikawa:2008iq}
that the non-linearity parameter reaches its maximal
value $f_{\rm NL}({\rm max})\sim \epsilon_*/\sqrt{\Gamma_\curv/m_\curv}$  for
$\lambda\sim 1$, i.e., for $\curv_*\sim
\sqrt{\Gamma_\curv/(m_\curv\epsilon_*) }\mP$. A significant non-Gaussianity is thus possible
if $\epsilon_*\gg \sqrt{\Gamma_\curv/m_\curv}$. 
It is easy to extend the above procedure  for the computation of the trispectrum~\cite{Sasaki:2006kq}.

Moreover, in the curvaton scenarios, isocurvature perturbations can be present. Even if their contribution to the power spectrum is constrained to be small,  they could contribute significantly to non-Gaussianities. It is thus interesting to study the non-Gaussianities of isocurvature perturbations as well (see e.g. \cite{Kawasaki:2008sn,Langlois:2008vk,Kawasaki:2008pa,Moroi:2008nn}). Non-Gaussianities in modulaton scenarios have also been investigated (see e.g. \cite{Zaldarriaga:2003my,Ichikawa:2008ne,Langlois:2009jp}).

\subsubsection{Multi-field DBI inflation}
Multi-field DBI inflation is another example  where non-Gaussianities have been investigated. In this case, the three-point correlation functions of the scalar fields are not negligible and they can be computed from the third order action, which is given, in the small sound speed limit, by~\cite{lrst08a,lrst08b}
\begin{eqnarray}
S_{(3)}&=&\int {\rm d}t\, {\rm d}^3x\,  \left\{ \frac{a^3}{2 c_s^5 \dot \s}\left[(\dot Q_{\s} )^3+c_s^2 \dot Q_{\s}  (\dot Q_{s} )^2\right]  \right.
 \cr
 && \left. 
 - \frac{a}{2 c_s^3 \dot \s}\left[ \dot Q_{\s} (\partial  Q_{\s} )^2 -c_s^2 \dot{Q_{\s} }(\partial Q_s)^2+2 c_s^2 \dot {Q_s}\partial Q_{\s} \partial Q_s)\right]
 \right\}
 \label{S3}
\end{eqnarray}
in terms of the  instantaneous adiabatic and entropic perturbations. 
The contribution from the scalar field three-point functions  to the coefficient $f_{\rm NL}$ is  
\beq
f_{NL}^{(3)}=-\frac{35}{108}\frac{1}{c_s^2}\frac{1}{1+T^2_{{\cal R} {\cal S}} }=-\frac{35}{108}\frac{1}{c_s^2} {\cos^2} \Theta \,
\label{f_NL3}
\eeq 
which is similar to the single-field DBI result~\cite{ast,Chen:2005fe},  but with a suppression due to  the transfer between the entropic and adiavatic modes. 

In the  trispectrum,  multi-field effects induce a shape of non-Gaussianities that differs from the single-field case~\cite{Mizuno:2009cv}. 
Moreover, multi-field DBI inflation could also produce a local non-Gaussianity in addition to the equilateral one (see \cite{RenauxPetel:2009sj} for an explicit illustration).

\section{Conclusions}

As these notes have tried to emphasize, inflation provides an attractive framework to describe the very early Universe and to account for the ``initial'' seeds of the cosmological perturbations, which we are able to observe today with increasing precision. In particular, the idea that the present structures in the Universe arose from the gravitational amplification of quantum vacuum fluctuations is especially appealing. 

At present, inflation is more a general framework  than a specific theory and there exists a plethora of models, based on various types of  motivation, which  can all satisfy the present observational data. 
The simplest models, based on a slow-rolling  single field, produce only adiabatic perturbations, with negligible non-Gaussianities, but with a possibly detectable amount of gravitational waves for the large-field subclass. 

More sophisticated models,  involving multiple scalar fields
or non-standard kinetic terms, can lead to a much richer spectrum of possibilities: isocurvature perturbations that could be correlated with the adiabatic ones, or a detectable level of non-Gaussianities.

Any clear evidence in the future  of one or several of these additional features (gravitional waves, isocurvature perturbations and/or primordial non-Gaussianities)  would allow us to discriminate between the main species of inflationary models and would thus have  a huge impact on our understanding of the early Universe.

\acknowledgement{I would like to thank the organizers of the Second TRR33 Winter School  for inviting me to a stimulating school. I am  also grateful to S\'ebastien Renaux-Petel for his useful comments on an earlier version of these notes. }

\end{document}